\documentclass[aps,prd,preprint,nofootinbib]{revtex4-1}
\usepackage{amsfonts}
\usepackage{mathrsfs}
\usepackage{graphicx}
\usepackage{amsmath}
\usepackage{amssymb}
\usepackage{epsfig}
\usepackage{graphicx}
\usepackage{slashed}
\usepackage{color}
\usepackage{multirow}
\usepackage{ulem}
\usepackage{adjustbox}

\usepackage{stmaryrd}
\newcommand{\bqa}{\begin{eqnarray}}
	\newcommand{\eqa}{\end{eqnarray}}
\newcommand{\beq}{\begin{equation}}
	\newcommand{\eeq}{\end{equation}}
\allowdisplaybreaks[1]

\graphicspath{{fig/}{dia/}} \DeclareGraphicsExtensions{.eps}
\begin{document}
	
	\title{Anatomy of $\Lambda_c^+$ semileptonic decays }

	\author {Chao-Qiang Geng\footnote{cqgeng@ucas.ac.cn}, Xiang-Nan Jin\footnote{jinxiangnan21@mails.ucas.ac.cn} and  Chia-Wei Liu\footnote{chiaweiliu@ucas.ac.cn}}
	\affiliation{	
		School of Fundamental Physics and Mathematical Sciences, Hangzhou Institute for Advanced Study, UCAS, Hangzhou 310024, China\\
		University of Chinese Academy of Sciences, 100190 Beijing, China
	}
	\date{\today}

	\begin{abstract}
We present a systematic study of $\Lambda_c^+ \to {\cal B}_q \ell^+ \nu_\ell $ with ${\cal B}_q = (\Lambda,  n)$ and $\ell =( e, \mu)$, examining all the possible decay observables based on the homogeneous bag model (HBM) and  lattice QCD (LQCD). With the HBM, we find that the branching fractions and  polarization asymmetries of the daughter baryon $\Lambda$ are ${\cal B}(\Lambda_c^+ \to \Lambda e^+ \nu _e, \Lambda \mu^+ \nu_\mu, n \ell ^+ \nu_\ell ) = (3.78 \pm 0.25, 3.67\pm 0.23, 0.40\pm 0.04 )\%$ and $\alpha_\Lambda (\Lambda_c^+ \to \Lambda e^+ \nu_e,\Lambda\mu^+ \nu_\mu ) =(-82.6,-82.3)\%$, respectively. From the LQCD, we obtain that $\alpha_{\Lambda}(\Lambda_c^+ \to \Lambda e^+ \nu_e,  \mu^+ \nu_\mu  ) = (-87.4\pm 1.0,-87.2\pm 1.0)\%$. We also explore the time-reversal asymmetries due to new physics beyond the standard model. All our results are consistent with the current experimental data, while some of them are accessible to the experiments at BESIII and Belle II.
	\end{abstract}

	\maketitle
	
	\section{Introductions}
	 
	Very recently,
	the BESIII~\cite{BESIII:2022ysa} collaboration  has reported the decay branching fraction of  $\Lambda_c^+ \to \Lambda e^+\nu_e$
	to be
	\begin{eqnarray}  
		{\cal B}(\Lambda_c^+ \to \Lambda e^+\nu_e) =(3.56\pm 0.11\pm 0.07)\%  \,,
	\end{eqnarray}
	which is the most precise measurement in the heavy baryon semileptonic decays up to date, providing an excellent opportunity to examine the standard model (SM) and various specific quark models. 
	On the other hand, the polarization asymmetry of the daughter baryon $\Lambda$ ($\alpha_\Lambda$) was measured 
	 nearly 20 years ago at CLEO~\cite{CLEO:2004txf}, given by
	\begin{eqnarray}
		\alpha_\Lambda(\Lambda_c^+ \to \Lambda e^+\nu_e) =(-0.86\pm 0.03\pm 0.02)\% \,.
	\end{eqnarray}
	 It is reasonable to expect that  some angular observables including $\alpha_\Lambda$ can be soon measured at BESIII and Belle II. 
	In the view of the recent experiments  of $\Lambda_c^+$ at BESIII~\cite{EXP1,EXP2,EXP3,EXP4,EXP5,BESIII:2016ffj},  the strange charmed baryons at Belle~\cite{Belle:2021crz,Belle:2021dgc,Belle:2022n,Belle:2022kqi}, and the doubly charmed baryons at LHCb~\cite{LHCb:2018pcs}, 
	it is no doubt that the era of the precision measurements in the charm baryons has begun. 	
	
	On the theoretical side,
the charmed baryons have also been considered with great interests  recently~\cite{TH0,TH1,TH2,TH3,TH4,TH5,TH6,TH7}. See Ref.~\cite{REV} for a review. 
In particular, due to the clean theoretical background,
 $ \Lambda_c^+ \to {\cal B}_q \ell^+ \nu_\ell$ with ${\cal B}_q = (\Lambda,  n)$ and $\ell = (e, \mu)$ are widely studied in the literature~\cite{Faustov:2019ddj,Gutsche:2015rrt}, including 
 the lattice QCD~(LQCD)~\cite{Meinel:2016dqj,Meinel:2017ggx,LATTICE1, LATTICE0},  
	light-front quark   model (LFQM)~\cite{Li:2021qod,Zhao:2018zcb,Geng:2020gjh}, and  $SU(3)$ flavor~$(SU(3)_F)$ symmetry~\cite{He:2021qnc,Geng:2019bfz}. However, there are some tensions. For instance, the branching fractions of $\Lambda_c^+ \to \Lambda \ell^+\nu_\ell$ based on the LFQM
  in Refs.~\cite{Zhao:2018zcb} and ~\cite{Li:2021qod,Geng:2020gjh} deviate by a factor of 2. Clearly, a different angle in analyzing $ \Lambda_c^+ \to {\cal B}_q \ell^+ \nu$ may shed light on the problem.  
  
	In this paper, we concentrate on the decays of  $\Lambda_c^+ \to {\cal B}_q \ell^+ \nu $ with the bag model (BM) and LQCD.
We  compute all the possible decay observables of $\Lambda_c^+ \to {\cal B}_q \ell^+ \nu $ in the SM and
discuss some of the possible effects from new physics (NP). In particular, we discuss the time-reversal~(T) asymmetries. As these asymmetries are contaminated little by the hadronic uncertainties,  they provide a reliable way to probe NP. 
	
	This work is organized as follows. 
	In Sec.~II, we give the baryon wave functions and helicity amplitudes from the homogeneous BM (HBM). In Sec.~III, we present the angular distributions and extract the physical observables in $\Lambda_c^+ \to {\cal B}_q \ell^+ \nu $. In Sec.~IV, we study the T asymmetries from NP. Finally, we conclude this study in Sec.~V.
	
	\section{Baryon wave functions and Helicity amplitudes}
	The amplitudes of $ \Lambda_c^+ \to  {\cal B}_q \ell^+ \nu$  are  given as 
	\begin{eqnarray}\label{eq1}
		&&\frac{G_F}{\sqrt 2} V_{cq}  g ^{\mu \nu}    \overline{v} {\gamma_{\mu}} (1-{\gamma_5}) u   \langle {\cal B}_q | \overline{q}{\gamma_{\nu}}(1-{\gamma_5})c | \Lambda _c \rangle,
	\end{eqnarray}
	where 
 $V_{cs}=0.973$ and $V_{cd}=0.221$ are the Cabibbo-Kobayashi-Maskawa quark mixing matrix elements~\cite{pdg}, 	$G_F$  is the Fermi constant, $q=(s,d)$ for ${\cal B}_q =(\Lambda,n)$, and $\overline{v}$ and $u$ are the Dirac spinors of $\nu$ and $\ell^+$, respectively. We  expand the Minkowski metric $g^{\mu\nu}$ by
	\begin{equation}\label{eq2}
		g^{\mu\nu} = \varepsilon_t ^{\mu } (q)\varepsilon_t^{\ast \nu} (q)- \sum_{\lambda = 0,\pm } \varepsilon^\mu _\lambda(q) \varepsilon^{\ast \nu}  _ \lambda (q) \,,  
	\end{equation}
	where  $\varepsilon$ and $q$ are the polarization vector and four-momentum of the off-shell $W$ boson~($W^*$), respectively, and 
	the subscript in $\varepsilon$ denotes the helicity of $W^*$.
	In the center of frames of $W^*$ and $\Lambda_c^+$, we have~\cite{Timereversal}
	\begin{equation}\label{restPolar}
		{\varepsilon} ^\mu _\pm  = \frac{1}{\sqrt{2}}(0,  \pm 1, i , 0 )^T\,, \quad {\varepsilon}_0^\mu  = (0 ,0,0,-1 )^T \,, \quad  {\varepsilon}^\mu  _t = ( - 1 ,0,0,0 )^T\,,
	\end{equation}
	and
	\begin{equation}\label{qframe} 
		{\varepsilon} ^\mu _\pm  = \frac{1}{\sqrt{2}}(0,  \mp 1, i , 0 )^T\,, \quad {\varepsilon}_0^\mu  = \frac{1}{\sqrt{q^2}}(-|\vec{q}\,| ,0,0,q^0 )^T \,, \quad  {\varepsilon}^\mu  _t =\frac{-1}{\sqrt{q^2}} q ^\mu ,
	\end{equation}
	respectively. We note that in the calculations, we always choose the 3-momentum of the outgoing  fermion toward  the $\hat{z}$ direction.
	
	By inserting Eq.~\eqref{eq2} into Eq.~\eqref{eq1}, the three-body problem is reduced to a product of the two-body ones, given by
	\begin{eqnarray}
		\frac{G_F}{\sqrt{2}}V_{cq} \left(
		L_tB_t - \sum_{\lambda= 0,\pm }  L_\lambda B_\lambda 
		\right) 
	\end{eqnarray}
	along with 
	\begin{eqnarray}\label{BL}
		&&L_{\lambda_W}  = \varepsilon_{\lambda_W}^{\mu} 
		\overline{v}_{\ell} {\gamma_{\mu}} (1-{\gamma_5}) u\,,~~~ B_{\lambda_W}  = \varepsilon_{\lambda_W}^{\ast \mu} 
		\langle {\cal B}_q |  \overline{s} {\gamma_{\mu}} (1-{\gamma_5})c |\Lambda_c^+ \rangle\,, 
	\end{eqnarray}
	describing  $W^{*} \to \ell^+ \nu $ and  $\Lambda_c^+ \to  {\cal B}_q W^ * $, respectively. 
	
	As $L_{\lambda_W}$ and  $B_{\lambda_W}$ are Lorentz scalars, they can be calculated in different Lorentz frames. 
	We adopt the rest frames of $W^*$ and  $\Lambda_c^+$ for $L_{\lambda_W}$ and 
	$B_{\lambda_W}$, respectively.  In the SM, the helicity amplitudes are   given as
	\begin{eqnarray}\label{spatial}
		&&h_{\pm }
		= \varepsilon_{-\frac{1}{2} \pm \frac{1}{2} } ^ {\mu }
		\overline{v}_- {\gamma_{\mu}} (1-{\gamma_5})u_{\pm } \,. \nonumber\\
		&&H_{\lambda_q \lambda_W }
		= \varepsilon_{\lambda_W} ^ {\ast \mu }\left( {\cal V}_\mu^{\lambda_q \lambda_c} -
		{\cal A}_\mu^{\lambda_q \lambda_c}\right) \,, 
	\end{eqnarray}
	with $\lambda_c = \lambda_q - \lambda_W$ and
	\begin{eqnarray}\label{spatial2}
		{\cal V}_\mu^{\lambda_q \lambda_c} & =&
		\langle  {\cal B}_q,  \hat{J}\cdot\hat{p} = \lambda_q, \vec{v} = v_0\hat{z}  |  \overline{q} {\gamma_{\mu}} c |\Lambda_c^+, J_z = \lambda_c , \vec{v} = 0     \rangle\nonumber\\
		&=&N_q\langle duq,  J^3_z = \lambda_q, \vec{v} = v_0\hat{z}  |  \overline{q}_3 {\gamma_{\mu}} c_3 |duc, J^3_z = \lambda_c , \vec{v} = 0     \rangle\,,\nonumber\\
		{\cal A}_\mu^{\lambda_q \lambda_c}& =&
		\langle {\cal B}_q,  \hat{J}\cdot\hat{p} = \lambda_q, \vec{v} = v_0\hat{z}  |  \overline{q} {\gamma_{\mu}} \gamma_5  c |\Lambda_c^+, J_z = \lambda_c , \vec{v} = 0     \rangle\nonumber\\
		&=&N_q \langle duq,  J^3_z = \lambda_q, \vec{v} = v_0\hat{z}   |  \overline{q}_3 {\gamma_{\mu}} \gamma_5 c_3 |duc, J^3_z = \lambda_c , \vec{v} = 0   \rangle\,,
	\end{eqnarray}
	where $N_q=(1, \sqrt{3/2})$ for $q=(s,d)$ are the spin-flavor overlappings,
	the subscript of Dirac spinors stands for helicity, 
	$\overline{s}_3$ and $c_3$  act only on the third quarks,  $J^3$ is the angular momentum of the third quark, and $\vec{v}$ represents the velocity. 
	
	The adopted convention of the Dirac spinors can be found in Appendix of Ref.~\cite{Timereversal}. 
	The  helicity amplitudes $h_\pm$ are calculated as
	\begin{equation}\label{ratiosof Lepton}
		h_+ = -  2 \sqrt{2 (q^2 - M_{\ell} ^2) } \,,~~~
		h _-  
		= \sqrt{\delta_\ell} h _ +\,,
	\end{equation}
	where $\delta_\ell= {2 M_\ell^2/q^2 }$ and $M_\ell$ corresponds to the charged lepton mass. Due to the left-handed nature of the weak interaction, we have $h_+\gg h_-$ as $\ell^+$ has a positive helicity in the massless limit.

	On the other hand, ${\cal V}$ and ${\cal A}$ depend on the baryon wave functions and vary  in quark  models. 
	The relevant baryon wave functions are given as 
	\begin{eqnarray}\label{wave}
		&&	|\Lambda_c^+, \updownarrow\rangle = \int\frac{1}{\sqrt{6} } \epsilon^{\alpha \beta \gamma} d _{a\alpha}^\dagger(\vec{x}_1) u_{b\beta}^\dagger(\vec{x}_2) c_{c\gamma}^\dagger (\vec{x}_3) \Psi_{A_\updownarrow(duc)}^{abc} (\vec{x}_1,\vec{x}_2,\vec{x}_3) [d^3  \vec{x}] | 0\rangle\,,\nonumber\\
		&&	|\Lambda, \updownarrow\rangle = \int\frac{1}{\sqrt{6} } \epsilon^{\alpha \beta \gamma} d _{a\alpha}^\dagger(\vec{x}_1) u_{b\beta}^\dagger(\vec{x}_2) s_{c\gamma}^\dagger (\vec{x}_3) \Psi_{A_\updownarrow(dus)}^{abc} (\vec{x}_1,\vec{x}_2,\vec{x}_3) [d^3  \vec{x}] | 0\rangle\,,\nonumber\\
		&&	|n, \updownarrow\rangle = \int\frac{1}{2\sqrt{3} } \epsilon^{\alpha \beta \gamma} d _{a\alpha}^\dagger(\vec{x}_1) u_{b\beta}^\dagger(\vec{x}_2) d_{c\gamma}^\dagger (\vec{x}_3) \Psi_{A_\updownarrow(dus)}^{abc} (\vec{x}_1,\vec{x}_2,\vec{x}_3) [d^3  \vec{x}] | 0\rangle\,,
	\end{eqnarray}
	where  $\epsilon^{\alpha \beta\gamma}$ is the totally antisymmetric tensor, 
	the Latin and Greek letters stand for the spinor and color indices, and $q^\dagger(\vec{x})$ is the creation operator of a  quark at $\vec{x}$ and  $t=0$. 
	Without specifying the   distributions of $\Psi$, the wave functions in Eq.~\eqref{wave} are  general results of quark models in the instant form.  In the MIT BM, the distributions read as~\cite{bag} 
	\begin{equation} \label{orginal}
		\begin{aligned}
			\Psi_{A_{\updownarrow}(q_1q_2q_3)}^{abc} ( \vec{x}_1, \vec{x}_2 ,  \vec{x}_3 ) =& \frac{{\cal N}}{\sqrt{2}} \left( \phi^a_{q_1\uparrow}(\vec{x}_1) \phi^b_{q_2\downarrow}(\vec{x}_2) - \phi^a_{q_1\downarrow}(\vec{x}_1) \phi^b_{q_2\uparrow}(\vec{x}_2)\right)  \phi^c_{q_3\updownarrow}(\vec{x}_3) ,
		\end{aligned}
	\end{equation}
	where ${\cal N}$ is the normalization constant, and $\phi^a_{q\updownarrow}(\vec{x}) $ is a bag state centering at $\vec{x}=0$ 
	\begin{equation}\label{quark_wave_function}
		\phi_{q \updownarrow}^a(\vec{x}) = \left(
		\begin{array}{c}
			\omega_{q}^+ j_0(p_{q}r) \chi_\updownarrow\\
			i\omega_{q}^- j_1(p_{q}r) \hat{x} \cdot \vec{\sigma} \chi_\updownarrow\\
		\end{array}
		\right)_a\,,
	\end{equation}
	with $\chi _\uparrow = (1,0)^T$ and $\chi_\downarrow = (0,1)^T$ representing $J_z = \pm 1/2$,  and $j_{0,1}$ the spherical Bessel functions,
	respectively. Here, the kinematic factors are defined as 
	$\omega_{q}^\pm  = \sqrt{E_{q}^k \pm m_{q}} $ with $E_{q}^k$, $p_{q}$ and $m_{q}$ the kinematic energy, 3-momentum and mass of the quark. In turn, $p_{q}$ has to obey the boundary condition
	\begin{equation}
		\tan (p_{q}R) = \frac{p_{q} R}{ 1 - m_{q} R - E_{q}^k R  }\,,
	\end{equation}
	with $R$ the bag radius. 
	
	Although the MIT BM successfully explains most of the low-lying baryon masses, it is problematic when it applies to decays. It is due to that the baryon wave functions in Eq.~\eqref{orginal} are localized at $\vec{x} = 0$. According to the Heisenberg uncertainty principle, it can not be a momentum eigenstate, which is a basic requirement in calculating   decay widths. To resolve the problem, we take the  baryon wave functions as  infinite  linear superpositions of 
	Eq.~\eqref{orginal}, given by
	\begin{equation} \label{xdelta}
		\begin{aligned}
			\Psi_{A_{\updownarrow}(q_1q_2q_3)}^{abc} ( \vec{x}_1, \vec{x}_2 ,  \vec{x}_3 ) =& \frac{{\cal N}_{{\cal B}}}{\sqrt{2}} \int \left( \phi^a_{q_1\uparrow}(\vec{x}_1- \vec{x}_\Delta) \phi^b_{q_2\downarrow}(\vec{x}_2-\vec{ x}_\Delta) \right. \\
			&\left. - \phi^a_{q_1\downarrow}(\vec{x}_1- \vec{x}_\Delta) \phi^b_{q_2\uparrow}(\vec{x}_2-\vec{ x}_\Delta) \right) \phi^c_{q_3\updownarrow}(\vec{x}_3- \vec{x}_\Delta) d^3 \vec{x}_\Delta,
		\end{aligned}
	\end{equation}
	which clearly distribute homogeneously over the space, and thus are named as the HBM~\cite{Liu:2022pdk}. 
	The overall normalization constants, depending on the quark components, are calculated as 
	\begin{equation}
		{\cal N}_{{\cal B}}=
		\frac{1}{\overline{u}_{{\cal B}} u_{{\cal B}} }\int d^3\vec{x}_\Delta\prod_{i=1,2,3}   D^0_{q_i} (\vec{x}_\Delta ) \,,
	\end{equation}
	with 
	\begin{equation}
		D^v_q (\vec{x}_\Delta ) = \frac{1 }{\gamma}
		\int d^3 \vec{x } \phi_q^\dagger\left(\vec{x}^+ \right)  \phi_q \left(\vec{x}^- \right)  e ^ { - 2 i E_q \vec{v}\cdot \vec{x}} \,,    
	\end{equation}
	where
	$\vec{x}^\pm = \vec{x} \pm \vec{x}_\Delta/2$, and 
	 the dependencies on $v$ are  for the later convenience.  
The wave functions of the low-lying  baryons are collected in Appendix A. 
	
	Conventionally, though not necessarily,  matrix elements are  evaluated at the Briet frame, of which the initial and final baryons have opposite velocities $-\vec{v}$ and $\vec{v}$, respectively. They are related  to Eq.~\eqref{spatial} as 
	\begin{equation}\label{afterboost}
		\begin{aligned}
			&{\cal V}_\mu^{\prime  \lambda_q \lambda_c} = \Lambda _\mu\,^\nu {\cal V}_\nu^{\lambda_q \lambda_c} \,,~~~ {\cal A}_{ \mu}^{\prime \lambda_q \lambda_c} = \Lambda_\mu\,^\nu {\cal A}_{\nu} ^{\lambda_q \lambda_c}   \,,
		\end{aligned}
	\end{equation}
	where $\Lambda^\mu\,_\nu$ is a Lorentz boost toward the $-\hat{z}$ direction, defined as $\Lambda^3\,_0= -\gamma v$ with  $\gamma = \sqrt{1/(1-v^2)}$. From the baryon wave functions given in Eq.~\eqref{xdelta}, we   obtain
	\begin{equation}
		\begin{aligned}\label{matr}
			{\cal V}_\mu^{\prime \lambda_q \lambda_c} = {\cal N}_{{\cal B}_c} {\cal N}_{{\cal B}_q} \int d^3\vec{x}_\Delta \Upsilon^{\lambda_q \lambda_c} _{\mu }(\vec{x}_\Delta) \prod_{l=d,u } D^{v}_{q}( \vec{x}_\Delta)\,,\\
			{\cal A}_\mu^{\prime  \lambda_q \lambda_c} = {\cal N}_{{\cal B}_c} {\cal N}_{{\cal B}_q} \int d^3\vec{x}_\Delta \Upsilon^{\lambda_q \lambda_c} _{5 \mu }(\vec{x}_\Delta) \prod_{l=u,d } D^{v}_{q}( \vec{x}_\Delta)\,,
		\end{aligned}
	\end{equation}
	with 
	\begin{eqnarray}\label{master}
		\Upsilon_{\mu}^{\lambda_q \lambda_c}(\vec{x}_\Delta )
		&=&\int  d^3\vec{x}  \phi _{q{\lambda_1}}^\dagger\left(\vec{x} ^+ \right) S_{v} \gamma_0 \gamma_\mu  S_{-v} \phi_{c{\lambda_2}}\left(\vec{x} ^- \right) e^{2i(E_{u} + E_{d})\vec{ v}\cdot \vec{ x}     }\,,\nonumber\\
		\Upsilon_{5 \mu}^{\lambda_q \lambda_c}(\vec{x}_\Delta)
		&=&\int  d^3\vec{x}  \phi _{q{\lambda_q}}^\dagger\left(\vec{x} ^+ \right) S_{v} \gamma_0 \gamma_\mu  \gamma_5  S_{-v} \phi_{c{\lambda_c}}\left(\vec{x} ^- \right) e^{2i(E_{u} + E_{d})\vec{ v} \cdot \vec{ x} }\,,
	\end{eqnarray}
	where  $S_{\pm v} = a_+ \pm  a_-\gamma^0\gamma^3$ are  the Lorentz boost matrix for Dirac spinors toward the $\pm \hat{z}$ direction, and $a_\pm = \sqrt{(\gamma \pm 1 )/2 }$. 
	In Eq.~\eqref{matr}, $\Upsilon_{(5)}$ describes the quark transition of the (axial) current operator, and $D_q^v$ are the overlappings of the spectator quarks. 
	
	The main uncertainties of the model calculations come from the quark energies.
	In this work,
	we take the values~\cite{Liu:2022pdk}
	\begin{eqnarray}
		&&M_p/3~\text{GeV}< E_u < 0.368~\text{GeV}\,,
	\end{eqnarray}
	with $M_p$  the proton mass.
We use the capital $M$ to represent hadron masses and the lower case $m$ for quark masses. The adopted bag radius and  quark masses are~\cite{Liu:2022pdk,bag}
	\begin{equation}
R= 4.9~\text{GeV}^{-1}\,,~~~		m_{u,d} = 0 \,,~~~m_s = (0.19\pm 0.09 )~\text{GeV}~~~  m_c = 1.655~\text{GeV}\,.
	\end{equation}
	By collecting Eqs.~\eqref{afterboost}, \eqref{matr} and \eqref{master}, one shall be able to evaluate Eq.~\eqref{spatial}, which completes the evaluations. See Ref.~\cite{Liu:2022pdk} for calculation details. 
	
	Before we end this section, we note that 
	the baryon transition matrix elements are conventionally parameterized as follows:
	\begin{eqnarray}\label{number45}
		&&\langle {\cal B}_q | \overline{s}{\gamma_{\nu}}(1-{\gamma_5})c | \Lambda ^+_c \rangle = \overline{u}_{q}\left[ \left(f _1(q^2) {\gamma_{\mu}} - i f _2(q^2) \frac{\sigma_{\mu \nu}}{M_c}q^{\nu} + f _3(q^2)\frac{q_{\mu}}{M_c} \right)\right.  \nonumber\\
		&&~~~~~~~~~~- \left. \left(g _1(q^2) {\gamma_{\mu}} -ig  _2(q^2) \frac{\sigma_{\mu \nu}}{M_c}q^{\nu} + g _3(q^2)\frac{q_{\mu}}{M_c} \right){\gamma_5}\right]  u_{c},
	\end{eqnarray}
	where $\overline{u}_q$ and $u_c$ are the Dirac spinors of ${\cal B}_q$ and $\Lambda_c^+$, $\sigma_{\mu \nu} = i(\gamma_\mu \gamma_{\nu} - \gamma_\nu \gamma_\mu)/2\,,$
	and $M_{q}$ and $M_c$ are the masses of ${\cal B}_q$ and $\Lambda_{c}^+$, respectively. 
	The form factors in Eq.~(\ref{number45}) can be numerically extracted  by matching Eqs.~\eqref{spatial} and \eqref{number45} once ${\cal V}$ and ${\cal A}$ are computed.  The relations between the  helicity amplitudes and form factors are 
	\begin{eqnarray}\label{helicity}
		&&	H_{\pm \frac{1}{2} \pm  1}=\sqrt{2 Q_{-}}\left(  f_1  +  \frac{M_{+}}{M_{c}} f_{2}\right)
		\pm  \sqrt{2 Q_{+}}\left(-g_1+ \frac{M_{-}}{M_{c}}g_2 \right)	\,,\nonumber\\
		&&	H_{\pm \frac{1}{2} 0} = - \sqrt{ \frac{Q_{-}}{q^{2}}} \left(M_{+} f_1 + \frac{q^{2}}{M_{c}} f_2 \right)\pm  
		\sqrt{ \frac{Q_{+}}{q^{2}}}\left(M_{-} g_1 - \frac{q^{2}}{M_{c}} g_2 \right)\,,
		\nonumber\\
		&&	H_{\pm \frac{1}{2} t}=- \sqrt{ \frac{Q_{+}}{q^{2}}}\left(M_{-} f_1 + \frac{q^{2}}{M_{c}} f_3  \right)\pm
		\sqrt{ \frac{Q_{-}}{q^{2}}}\left(M_{+} g_1 - \frac{q^{2}}{M_{c}} g_3 \right)\,, 
	\end{eqnarray}
	with $M_\pm = M_c \pm M_q$ and  $Q_\pm = M_\pm ^2 - q^2$.

	\section{Decay observables }
	
	In this section, we study  physical observables by the means of the angular distributions.
	For convenience, we take the following abbreviations:
	\begin{equation}\label{abbriev for tab}
		a_\pm = H_{\pm\frac{1
			}{2} 0 }\,,~~~b_\pm = H_{\mp\frac{1
			}{2} \mp 1} \,,~~~t_\pm = H_{ \pm \frac{1
			}{2} t} \,,~~~|\xi ^2| = |\xi_+^2| +|\xi_-^2|  \,,~~~|\xi _\Delta ^2| = |\xi_+^2| - |\xi_-^2|\,,
	\end{equation}
	with $\xi = a,b$ and $t$. 
	Here, we do not explicitly write down the $q^2$ dependencies of the helicity amplitudes.

	\subsection{Observables in $\Lambda_c^+ \to \Lambda (\to p \pi^- )\ell^+ \nu_\ell $}
	We start with  the partial decay width, given  as
	\begin{eqnarray}\label{number}
		&& \frac{\partial \Gamma^\ell }{\partial q^2}= 
		\zeta \big [\left(1 + \delta_\ell\right)\left (|a^2| + |b^2| \right)+ 3\delta_\ell |t^2|\big ]\nonumber\\
		&&\zeta = \frac{G_F^2 V_{cs}^2}{192\pi^3}\frac{(q^2-M_{\ell}^2)^2}{M_{ c }^2q^2}|\vec{p}_s|\,,
	\end{eqnarray}
	where $\vec{p}_s$ is the three-momentum of $\Lambda$ in the rest frame of $\Lambda_c^+$. 
	The angular momentum of a timelike $W^*$ in its rest frame is essentially zero, resulting in that 
	$\ell^+$ and $\nu_\ell $  have opposite helicities.
	As $t_\pm$ attribute to the timelike $W^*$ solely, they are always followed by $\delta_\ell$ as shown in Eq.~\eqref{number}. 
	Numerically,
	$\delta_\ell $ can be  taken as zero. 
	
	The branching fractions along with  those from the literature and  experiments   are collected in Table.~\ref{BRTable}, where
	Ref.~\cite{Gutsche:2015rrt} employs the covariant quark model (CQM), 
	Ref.~\cite{Faustov:2019ddj} considers the relativistic quark model (RQM),
	Ref.~\cite{Meinel:2016dqj} calculates the form factors by the LQCD, Refs.~\cite{Zhao:2018zcb,Geng:2020gjh, Li:2021qod} adopt the LFQM, and Refs.~\cite{He:2021qnc,Geng:2019bfz} utilize the $SU(3)_F$  symmetry. All the results in the literature show that  ${\cal B}(\Lambda_c^+ \to \Lambda e^+ \nu_e)$ is larger than 
	${\cal B}(\Lambda_c^+ \to \Lambda \mu^+ \nu_\mu)$. Therefore, an opposite behavior in the experimental measurements shall be a clear evidence of NP. The branching fractions in the literature are all in the same order as the  experimental ones. 
	The results of  $SU(3)_F$ are well consistent with those from the experiments.
	On the other hand, the branching fraction of $\Lambda_c^+ \to \Lambda e^+\nu_e$ in Ref.~\cite{Zhao:2018zcb}  from the LFQM 
	 is twice smaller than the experimental one. By adopting a different set of parameter input, Refs.~\cite{Geng:2020gjh, Li:2021qod} show that the LFQM is capable of explaining the experimental data, but the predictive power  is questionable in turn.  A parameter-independent study of the LFQM is clearly required.
	One of the great advantage of the BM is that the parameters are fitted from the mass spectra.
	In particular, the uncertainties are considerably smaller than other quark models. Our central values of the branching fractions are slightly larger than the ones from the experiments but smaller than those of the LQCD. 
	
	\begin{table}[b]
		\caption{   The branching fractions of $\Lambda_c^+ \to \Lambda \ell^+ \nu_\ell $ in units of $\%$, where the numbers in the parentheses are  the uncertainties counting backward in digits, {\it i.e. $3.75(25)=3.75\pm0.25$. }}
		\label{BRTable}
		\begin{tabular}{l|cccc}
			\hline
			\hline
			Model &\multicolumn{1}{c}{${\cal B}(\Lambda_c^+ \to \Lambda e^+\nu_e) $}
			&\multicolumn{1}{c}{${\cal B}(\Lambda_c^+ \to \Lambda {\mu}^+\nu_{\mu}) $ }\\
			\hline
		HBM&$ 3.78 (25)  $&$ 3.67 (23) $\\
			Data~\cite{BESIII:2022ysa,BESIII:2016ffj} &$ 3.56 (13) $&$ 3.49 (53)$\\
			CQM~\cite{Gutsche:2015rrt}&$ 2.78$&$2.69$\\
				RQM~\cite{Faustov:2019ddj}&$ 3.25 $&$3.14 $\\
			LQCD~\cite{Meinel:2016dqj} &$ 3.80 (22) $&$ 3.69 (22) $\\
			LFQM~\cite{Zhao:2018zcb}&$ 1.63 $&$ - $\\
			LFQM~\cite{Geng:2020gjh} &$ 3.55 (104) $&$ 3.40 (102) $\\
			LFQM~\cite{Li:2021qod} &$ 4.04 (75) $&$ 3.90(73) $\\
				$SU(3)_F$~\cite{Geng:2019bfz} & $3.6 (4) $ & $3.5(4)$\\
			$SU(3)_F$~\cite{He:2021qnc} &$ 3.62 (32) $&$ 3.45 (30) $\\
			\hline
			\hline
			
		\end{tabular}
	\end{table}

\begin{figure}[b]
		\includegraphics[width=.8\textwidth]{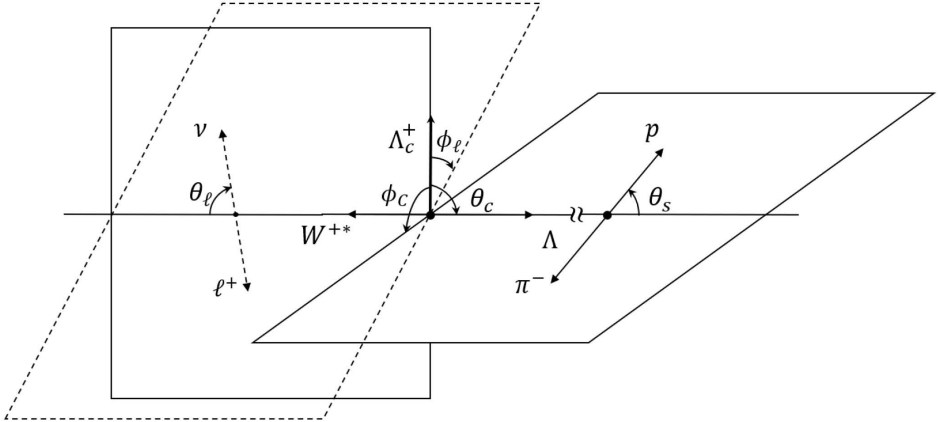}  
		\caption{ Definitions of angles, where the spin of $\Lambda_c^+$ points toward the direction of $\theta_c=0$.  } 
		\label{Baryon}
	\end{figure}

	To examine the decays further, we study the angular distributions of $\Lambda_c^+ \to \Lambda(\to p \pi^-) \ell^+ \nu_\ell $. 
	The fivefold angular  distributions  are obtained by piling up the Wigner-$d$ matrices of $d^J$, given as 
	\begin{eqnarray}\label{distru}
		&&{\cal D}(q^2, \vec{\Omega} ) \equiv \frac{\partial^6 \Gamma^\ell }{\partial q ^2 \partial \partial \vec{\Omega}  } = {\cal B}(\Lambda \to p \pi^+)  \frac{\zeta }{32\pi^2 } \sum_{\lambda_\ell \,,\lambda_p \,,\lambda_c  }\rho_{\lambda_c\lambda_c }\left| A_{\lambda_p } h_{ \lambda_\ell }\right|^2  \nonumber\\
		&&   \left|
		\sum_{\lambda_s\,, \lambda_W } (-1) ^{J_W  }
		H_{\lambda_q \lambda_W }
		d^{\frac{1}{2}}(\theta_c)^{\lambda_c }\, _{\lambda_s - \lambda_W }
		d^{\frac{1}{2}}(\theta_s)^{\lambda_s}\,_{\lambda_p }
		d^{J_W }(\theta_\ell )^{\lambda_W }\,_{ \frac{1
			}{2}- \lambda_\ell   }
		e^{i(\lambda_s \phi_c +\lambda_\ell  \phi_\ell )} 
		\right|^2,
	\end{eqnarray}
	where $\vec{\Omega} = (\cos \theta_c,  \cos \theta_s , \cos \theta_\ell  , \phi_c ,\phi_\ell)$, 
	$\rho_{\pm\pm } = (1\pm P_b)/2$ with $P_b$ the polarization fraction of $\Lambda_c^+$, 
	$\lambda_{c,s,\ell ,p} = \pm 1/2$ , 
	$|\vec{p}_c|  = \sqrt{Q_+Q_-}/2M_{c}$,  $A_\pm^2 = (1\pm \alpha)/2 $  with $\alpha$ the up-down asymmetry of $\Lambda \to p \pi^+$, and $J_W$ is  the  angular momentum of $W^*$. 
	The derivation can be found in Appendix of Ref.~\cite{Timereversal}, and the angles $\vec{\Omega}$ are defined in FIG.~\ref{Baryon}.
	By expanding Eq.~\eqref{distru}, we arrive at
	\begin{eqnarray}\label{exp}
		&& {\cal D}(q^2, \vec{\Omega})\propto 1 +  {\cal X}_2P_2+ {\cal X}_3 \cos \theta_\ell+ P_b \Big ( {\cal X}_4 \cos \theta_c + {\cal X}_5  \cos \theta_c P_2 + {\cal X}_6  \cos \theta_c \cos \theta_\ell \nonumber \\ 
		&& + \text{Re}( {\cal X}_7 e^{i \phi_{\ell}})\sin \theta_c \sin \theta_\ell + \text{Re} ({\cal X}_8e^{i \phi_{\ell}}) \sin \theta_c \sin \theta_\ell \cos \theta_\ell \Big )+ \alpha \Big ( {\cal X}_9\cos \theta_s  + {\cal X}_{10}\cos \theta_s P_2 \nonumber\\
		&&+ {\cal X}_{11}\cos \theta_s \cos \theta_\ell+ \text{Re} ({\cal X}_{12}e^{i\Phi}) \sin \theta_s \cos \theta_s + \text{Re}({\cal X}_{13}e^{i\Phi}) \sin \theta_s \cos \theta_s \cos \theta_\ell \Big) \nonumber  \\
		&& + P_b \alpha \Big(  {\cal X}_{14} \sin \theta_c \sin \theta_s P_2+ {\cal X}_{15} \cos \theta _c \cos \theta_s P_2+ {\cal X}_{16} \cos \theta_c \cos \theta_s + {\cal X}_{17} \cos \theta_c \cos \theta_s \cos \theta_\ell \nonumber\\
		&&+ \text{Re}( {\cal X}_{18}e^{-i \phi_c} ) \sin \theta _c \sin \theta_s P_2 +\text{Re}({\cal X}_{19}e^{i(\phi_c + 2\phi_{\ell})}) \sin \theta_c \sin \theta_s P_2+ \text{Re}({\cal X}_{20} e^{i \phi_c} ) \sin \theta_c \sin \theta_s \nonumber\\
		&&+\text{Re} ({\cal X}_{21} e^{i(\phi_c + 2\phi_{\ell})}) \sin \theta_c \sin \theta_s+ \text{Re}( {\cal X}_{22}(e^{i \phi_c}) \sin \theta _c\sin \theta_s \cos \theta_\ell
		\nonumber\\
		&&+ \text{Re}({\cal X}_{23} e^{i \phi_{\ell}} )\sin \theta_c \sin \theta_\ell \cos \theta_s + \text{Re}( {\cal X}_{24} e^{i\Phi} )\sin \theta_s \sin \theta_\ell \cos \theta_c\nonumber\\
		&&+ \text{Re}( {\cal X}_{25} e^{i \phi_{\ell}}) \sin \theta_c \sin \theta_\ell \cos \theta_s \cos \theta_\ell + \text{Re}( {\cal X}_{26}e^{i\Phi}) \sin \theta_s \sin \theta_\ell \cos \theta_c \cos \theta_\ell\Big ) \,.
	\end{eqnarray}
	where 
	$P_2 = (3 \cos\theta_\ell -1 )/2 $, $\Phi = \phi_c + \phi_\ell $, and  the observables ${\cal X}_{2\text{-}26}$ are defined in TABLE~\ref{table2}.
	Note that within the SM, $\xi_\pm$ are real, leading to  Re$({\cal X})={\cal X}$.  To obtain the distributions of  the charge conjugate processes, one can take the transformation of $\theta_\ell \to \pi-\theta_\ell$ and $\alpha \to \overline{\alpha}$ so that $ {\cal X}_i = \overline{ {\cal X}} _i$ in the absence of NP, where the overline denotes the charge conjugation.

	To test the results with the experiments, we  define the $q^2$ averages of the decay observables in TABLE~\ref{table2}, as  
	\begin{equation}
		\langle
		{\cal X}_i(\ell) 
		\rangle = 
		\frac{1}{\Gamma^\ell } \int_{M_\ell^2}^{M_-^2} \zeta {\cal X}_i d q^2.
	\end{equation}
The form factors calculated in the LQCD can be found in Refs.~\cite{Meinel:2016dqj,Meinel:2017ggx}, where  ${\cal B}$ are also provided. 
In this work, we calculate the angular observables from both the HBM and LQCD, listed in TABLE~\ref{table2}. 
The central values and uncertainties of the LQCD come from the nominal and higher order fits, respectively~\cite{Meinel:2016dqj,Meinel:2017ggx}. 
	Large parts of the uncertainties are canceled due to the correlations between  ${\cal X}_i$ and $\Gamma$.
	It is worth to mention that the HBM results agree well with the LQCD ones except for $\langle {\cal X}_{14,18,20}(\ell)\rangle$, attributed by  $a_+ a_-^*$.

	\begin{table}[!h]
		\caption{ Definitions of ${\cal X}_i$ and their average values in units of $\%$. }
		\label{table2}
		\begin{tabular}{l|l|rrrrrr}
			\hline
			\hline
			\multirow{2}{*}{$i$}&\multirow{2}{*}{ {\centering ${\cal X}_i(\Lambda_c^+ \to \Lambda \ell^+ \nu) $  }}
			&\multicolumn{2}{c}{HBM}
			&\multicolumn{2}{c}{LQCD}\\
			&&$\langle {\cal X}_i(e)  \rangle$&$\langle {\cal X}_i(\mu )  \rangle$&$\langle {\cal X}_i(e)  \rangle$&$\langle {\cal X}_i(\mu )  \rangle$ \\
			\hline$2~$ & $(2\delta_\ell-1)(|a|^2 - \frac{1}{2} |b|^2)$  & $-43.1(9)$& $-37.1(7)$& $-43.8(13)$& $-37.6(11)$& \\
			$3$ & $ - 6\delta_\ell\left(\text{Re}\left( a_+t_+^*\right) +\text{Re} \left( a_-t_-^*\right) \right )+ \frac{3}{2}|b_\Delta|^2$  & $35.2(10)$& $28.6(12)$& $40.2(12)$& $33.7(13)$& \\
			$4$ & $(\delta_\ell + 1)(|a_\Delta|^2 + |b_\Delta|^2) + 3\delta_\ell |t_\Delta|^2$  & $-35.7(13)$& $-35.4(13)$& $-33.7(15)$& $-33.3(15)$& \\
			$5$ & $(2\delta_\ell-1)(|a_\Delta|^2 - \frac{1}{2}|b_\Delta|^2)$  & $70.8(3)$& $64.0(1)$& $73.9(9)$& $67.0(8)$& \\
			$6$ & $- 6\delta_\ell \left[\text{Re}(a_+t_+^*) - \text{Re}(a_-t_-^*)\right] - \frac{3}{2}|b_\Delta|^2$  & $-35.2(10)$& $-28.5(12)$& $-40.2(12)$& $-33.7(13)$& \\
			$7$ & $\frac{3}{\sqrt{2}}\left[ 2 \delta_\ell (t_-b_+^* - b_-t_+^*) + (a_-b_+^* + b_-a_+^*)\right]$  & $-91.1(7)$& $-91.3(6)$& $-91.5(8)$& $-91.7(7)$& \\
			$8$ & $ \frac{3}{\sqrt{2}} (2\delta_\ell- 1) (b_-a_+^* - a_-b_+^*) $  & $-78.0(5)$& $-73.9(6)$& $-83.1(10)$& $-79.0(9)$& \\
			$9$ & $(\delta_\ell + 1)(|a_\Delta|^2 - |b_\Delta|^2) + 3\delta_\ell |t_\Delta|^2$  & $-82.6(0)$& $-82.3(0)$& $-87.4(10)$& $-87.2(10)$& \\
			$10$ & $ (2\delta_\ell-1)(|a_\Delta|^2 + \frac{1}{2}|b_\Delta|^2)$  & $47.4(10)$& $41.3(8)$& $47.1(12)$& $40.9(11)$& \\
			$11$ & $6\delta_\ell\left( \text{Re} \left( a_-t_-^*\right) - \text{Re}\left( a_+t_+^*\right)  \right ) - \frac{3}{2}|b|^2$  & $-56.9(9)$& $-50.2(11)$& $-56.2(13)$& $-49.8(14)$& \\
			$12$ & $ \frac{3 }{\sqrt{2}}\left[ 2 \delta_\ell (b_-t_-^* - t_+b_+^*) -  (a_+b_+^* + b_-a_-^*)\right]$  & $-49.0(1)$& $-47.5(0)$& $-38.8(25)$& $-38.0(24)$& \\
			$13$ & $\frac{3}{\sqrt{2}}\left[ 2 \delta_\ell (a_+b_+^*- b_-a_-^* ) + (b_-a_-^*- a_+b_+^* )\right]$  & $25.1(6)$& $23.8(5)$& $22.9(15)$& $21.8(14)$& \\
			$14$ & $ (2\delta_\ell+1)a_+ a_-^* $  & $-5.8(1)$& $-6.0(1)$& $-3.0(9)$& $-3.3(9)$& \\
			$15$ & $2\delta_\ell (|a|^2 + |b|^2)-(|a|^2 + 1/2|b|^2)$  & $-81.0(3)$& $-73.8(1)$& $-81.3(4)$& $-74.0(3)$& \\
			$16$ & $\delta_\ell(|a|^2 - |b|^2 + 3|t|^2 )+(|a|^2 - |b|^2)$  & $24.2(12)$& $24.0(12)$& $25.0(17)$& $24.5(17)$& \\
			$17$ & $-6\delta_\ell \left( \text{Re}(a_+t_+^*) +\text{Re}(a_-t_-^*)\right)  - \frac{3}{2}|b_\Delta|^2$  & $-35.2(10)$& $-41.0(8)$& $-40.3(12)$& $-46.3(12)$& \\
			$18$ & $(2-4\delta_\ell) a_- a_+^*$  & $-11.6(1)$& $-11.7(1)$& $-6.1(19)$& $-6.7(16)$& \\
			$19$ & $(1-2\delta_\ell) b_- b_+^*$  & $-14.5(1)$& $-14.2(1)$& $-12.5(6)$& $-12.3(6)$& \\
			$20$ & $-2\delta_\ell( a_+ a_-^* + 3t_+ t_-^*)-2a_+ a_-^*$  & $11.6(1)$& $11.6(1)$& $6.1(19)$& $6.4(17)$& \\
			$21$ & $(2\delta_\ell-1)b_-b_+^* $  & $14.5(1)$& $14.2(1)$& $12.5(6)$& $12.3(6)$& \\
			$22$ & $6 \delta_\ell(a_+t_-^* + t_+a_-^*)$  & $0.0(0)$& $0.4(0)$& $0.0(0)$& $0.5(2)$& \\
			$23$ & $\frac{3}{\sqrt{2}} \left[ (b_-a_+^* - a_-b_+^*)-2\delta_\ell (b_-t_+^* + t_-b_+^*)\right]$  & $78.0(5)$& $77.8(5)$& $83.1(10)$& $83.1(10)$& \\
			$24$ & $\frac{3}{\sqrt{2}} \left[ (b_-a_-^* - a_+b_+^*)-2 \delta_\ell (b_-t_-^* + t_+b_+^*)\right]$  & $25.1(6)$& $23.9(5)$& $22.9(15)$& $22.0(15)$& \\
			$25$ & $\frac{3}{\sqrt{2}}\left[2 \delta_\ell (a_-b_+^* + b_-a_+^*) - (a_-b_+^* + b_-a_+^*)\right]~~$  & $-49.0(1)$& $-47.7(0)$& $-38.8(25)$& $-38.1(24)$& \\
			$26$ & $\frac{3}{\sqrt{2}} \left[2 \delta_\ell (b_-a_-^*- a_+b_+^* ) - ( a_+b_+^* + b_-a_-^*)\right]$  & $-49.0(1)$& $-48.0(0)$& $-38.8(25)$& $-38.2(24)$& \\
			\hline
			\hline
		\end{tabular}
	\end{table}

\clearpage

\begin{figure}
	\includegraphics[width=.3\textwidth]{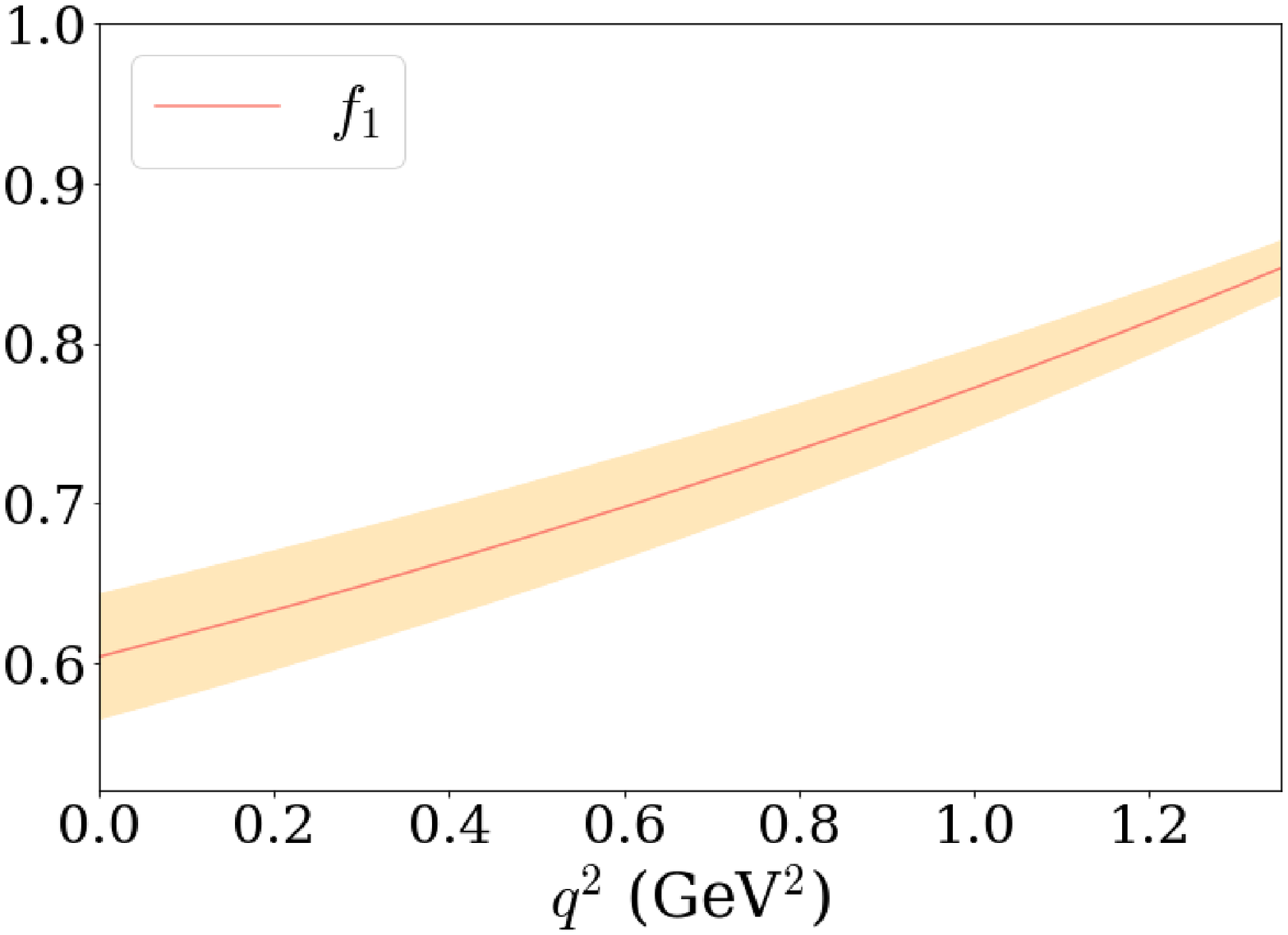}  
	\includegraphics[width=.3\textwidth]{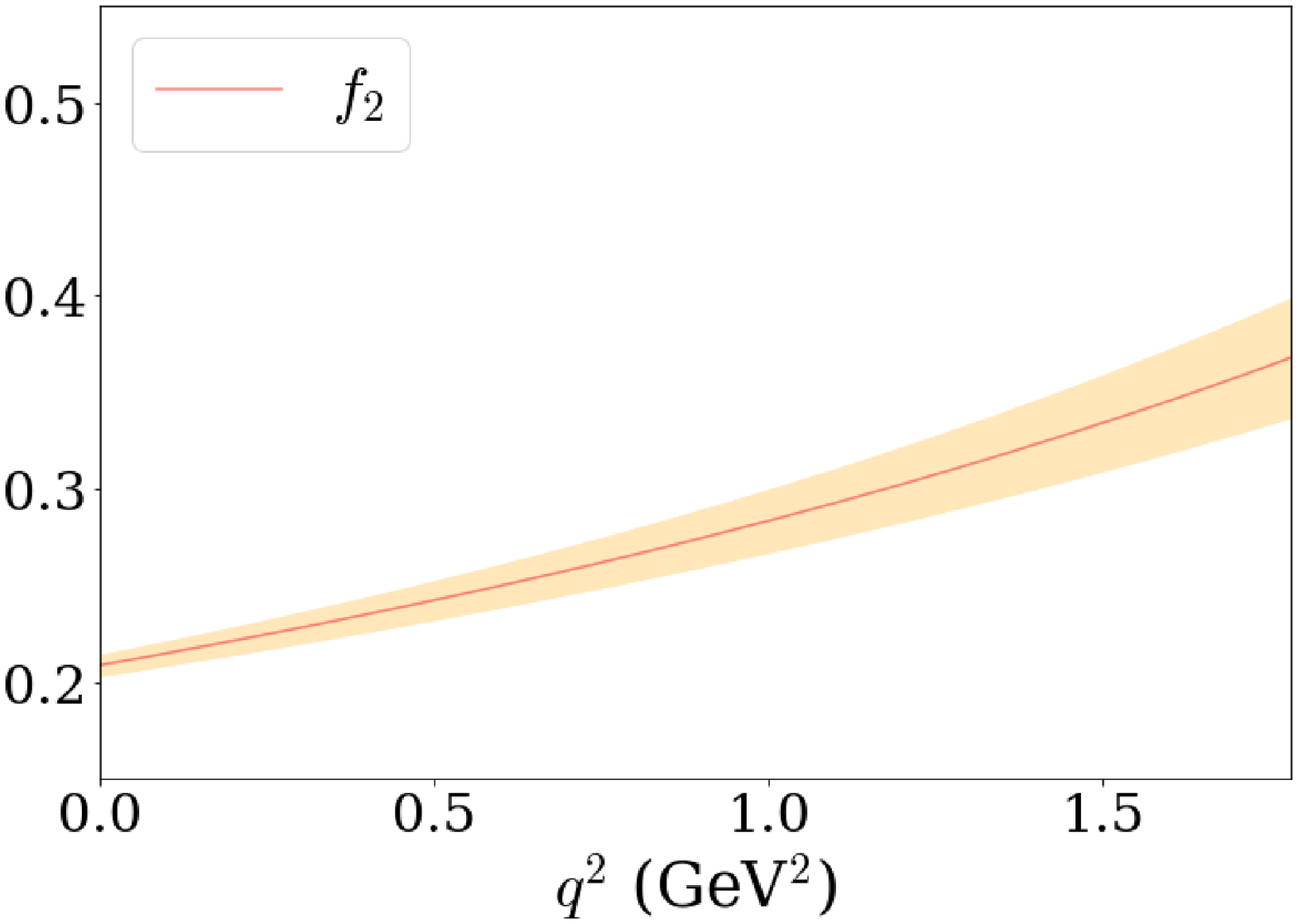} 
	\includegraphics[width=.3\textwidth]{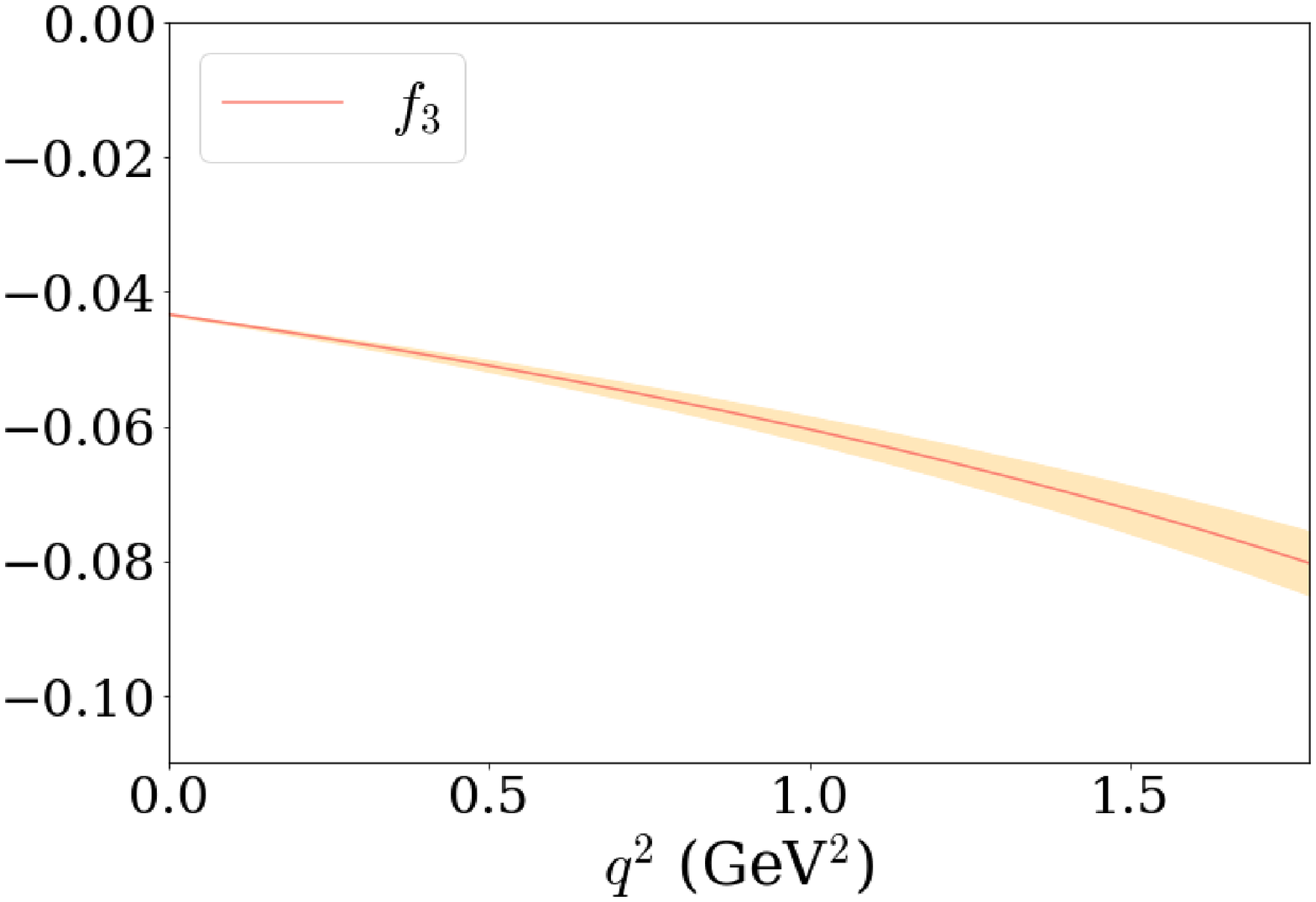} 
	\includegraphics[width=.3\textwidth]{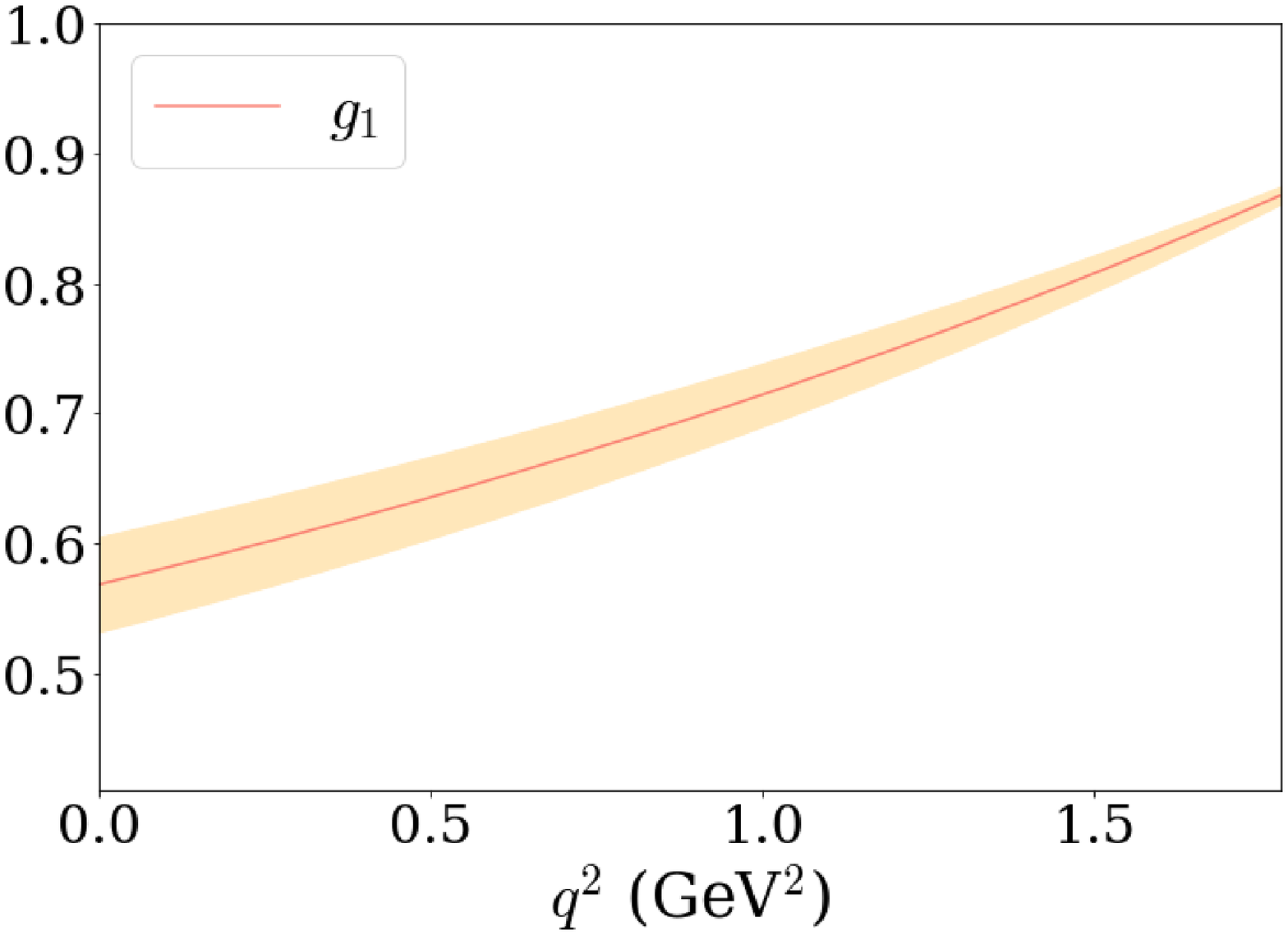}
	\includegraphics[width=.3\textwidth]{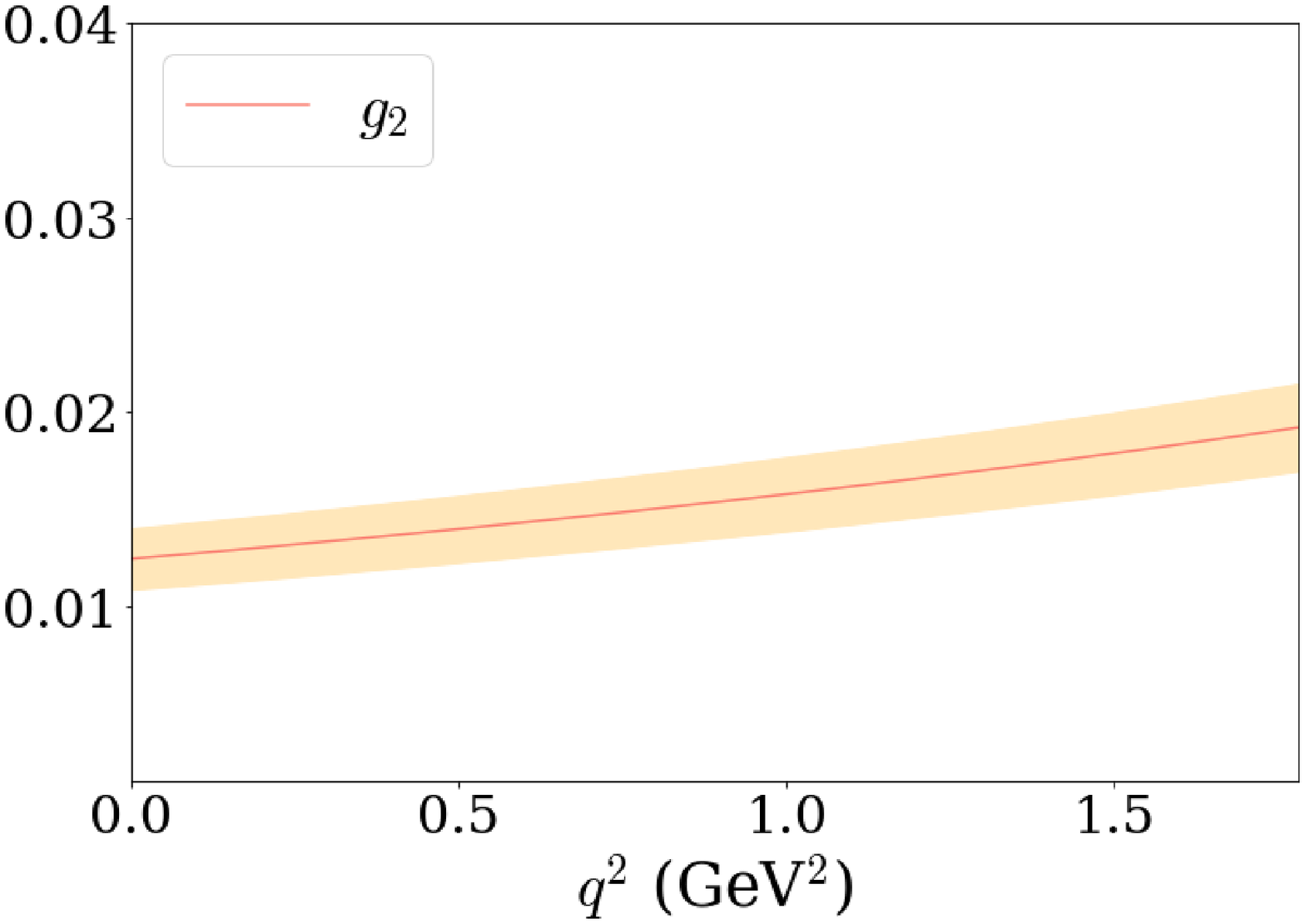} 
	\includegraphics[width=.3\textwidth]{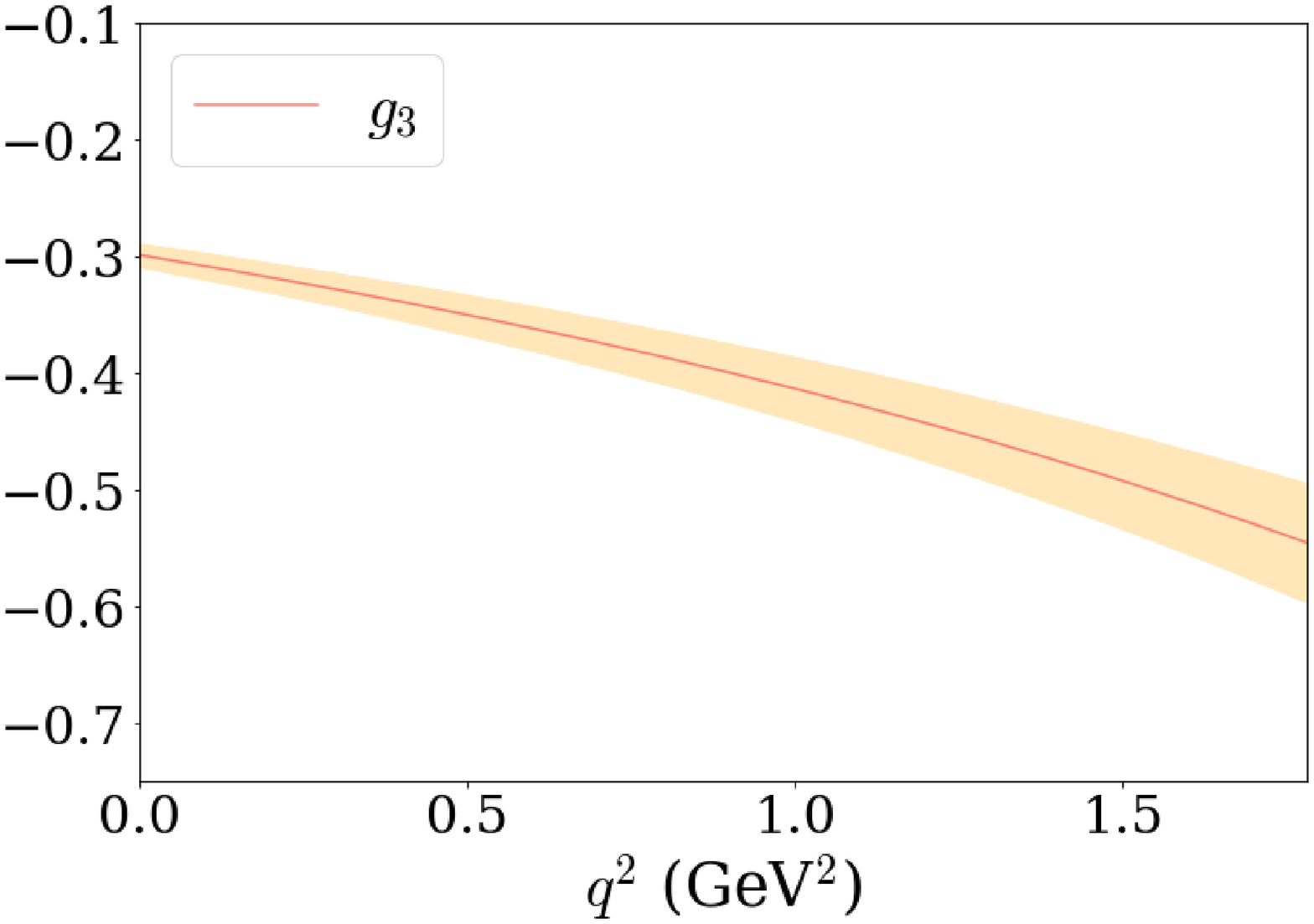}
	\caption{ The $q^2$ dependencies of the form factors in $\Lambda_c^+ \to \Lambda$ from the HBM, where the bands represent the uncertainties from the model calculations. } 
	\label{formfactorsLfigure}
\end{figure}

The reason of the deviations can be traced back to the form factors in the HBM, plotted in FIG.~\ref{formfactorsLfigure} and TABLE~\ref{formfTable}. 
By taking reasonable approximations of $f_1 = g_1$ and $g_2 =0$, we have $a_+ \propto f_2$ from  Eq.~\eqref{helicity}. Our approach  based on the BM  underestimates $f_2$ by a factor of two third, leading to a smaller $a_+$. This factor affects little on the branching fractions, but plays  important roles in some of the decay observables. 
We believe that it is somewhat a universal factor in the BM. For instance, the magnetic dipole moments of the octet baryons,  proportional to $f_2$, are also systematically underestimated  by a factor of two third in the MIT BM~\cite{bag}. 

\begin{table}
	\caption{ The form factors of $\Lambda_c^+\to \Lambda$ at $q^2=0$.}
	\label{formfTable}
	\begin{tabular}{lcccccccc}
		\hline
		\hline
		&\multicolumn{1}{c}{$f_1(0) $ }
		&\multicolumn{1}{c}{$f_2(0)$}
		&\multicolumn{1}{c}{$g_1(0)$ }
		&\multicolumn{1}{c}{$g_2(0)$ }
		\\
		\hline 
		HBM&$ 0.604(39)  $&$ 0.209 (5) $&$ 0.566(34) $&$ 0.012 (2) $\\
		LQCD~\cite{Meinel:2016dqj} &$ 0.643(23 )$&$0.308(36)  $&$0.572(15) $&$0.001(45) $\\
		\hline 
		\hline 
	\end{tabular}
\end{table}

To compare our results with the experiments, we calculate the form factors $f\in \{ g_\perp, g_+, f_\perp , f_+\} $ adopted by the experiments. The definitions of $f$ can be found explicitly in Ref.~\cite{Meinel:2017ggx}, while their $q^2$ dependencies are governed by 
\begin{equation}
f\left(q^2\right)=\frac{a_0^f}{1-q^2 /\left(m_{\text {pole }}^f\right)^2}\left[1+\alpha_1^f \times z\left(q^2\right)\right]\,,
\end{equation}
where $m_{\text {pole }}^f$ is the pole mass, $z = (\sqrt{t_+ - q^2 } - \sqrt{t_+-t_0} )/ (\sqrt{t_+ - q^2 } + \sqrt{t_+ - t_0}  )$ with $t_0 = (M_c-M_s)^2$, and $t_+ = (M_D + M_K)^2$.  The adopted masses are $(M_D , M_K , m_{\text{pole}}^{f_\perp, f_+}, m_{\text{pole}}^{g_\perp, g_+} )= (1.87 , 0.494, 2.112, 2.46)$~GeV. 

\begin{figure}[b]
	\includegraphics[width=.4\textwidth]{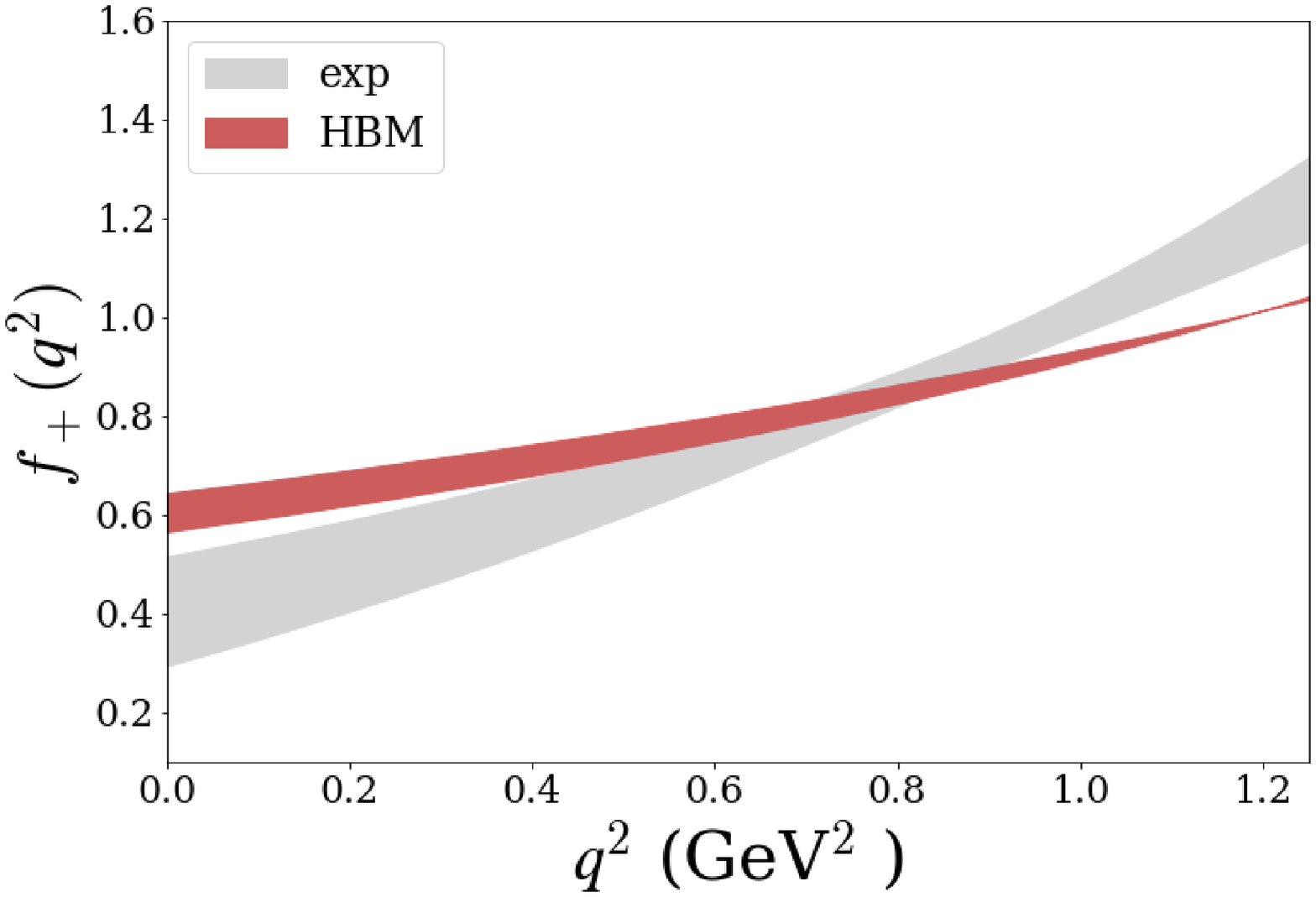}  
	\includegraphics[width=.4\textwidth]{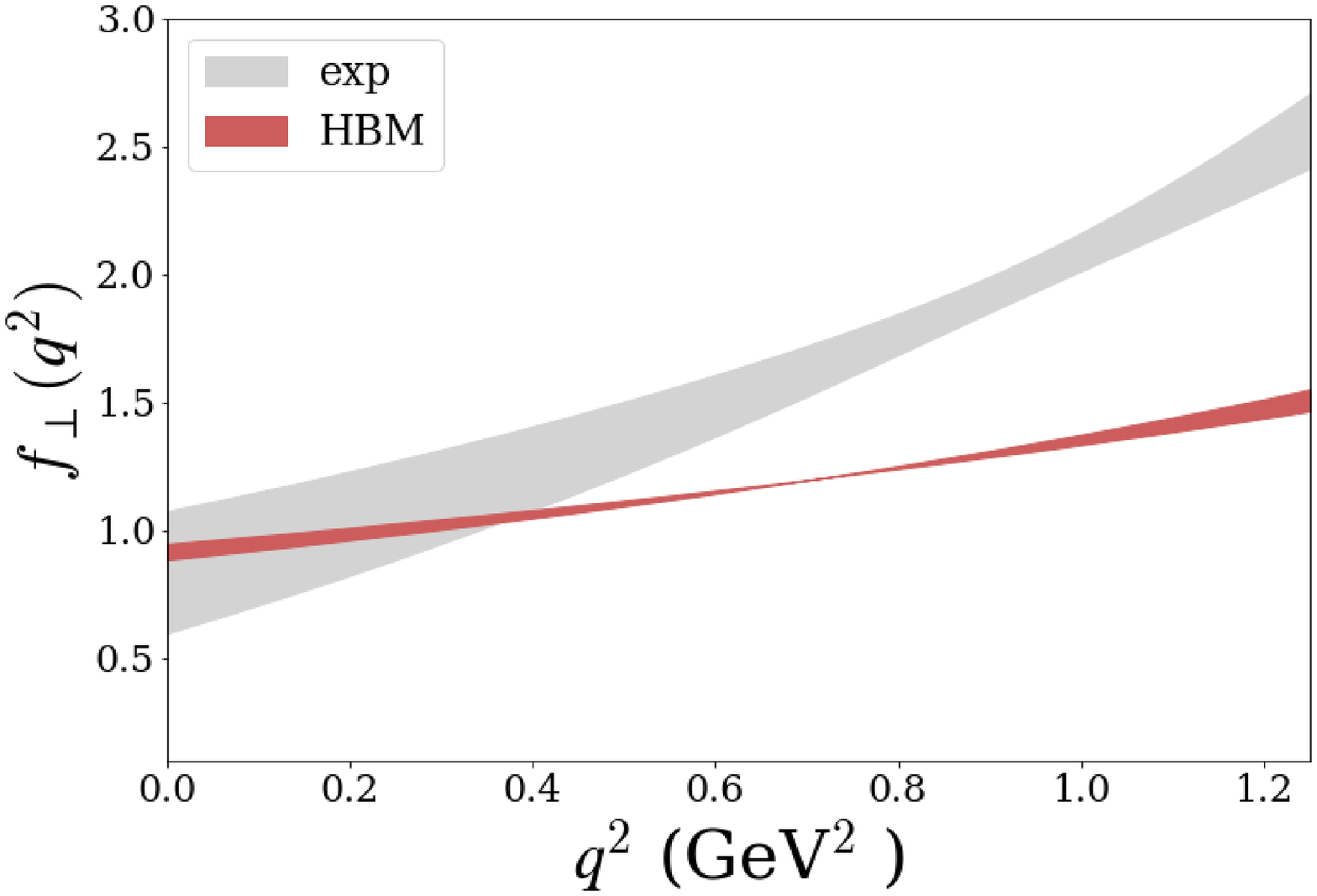} 
	\includegraphics[width=.4\textwidth]{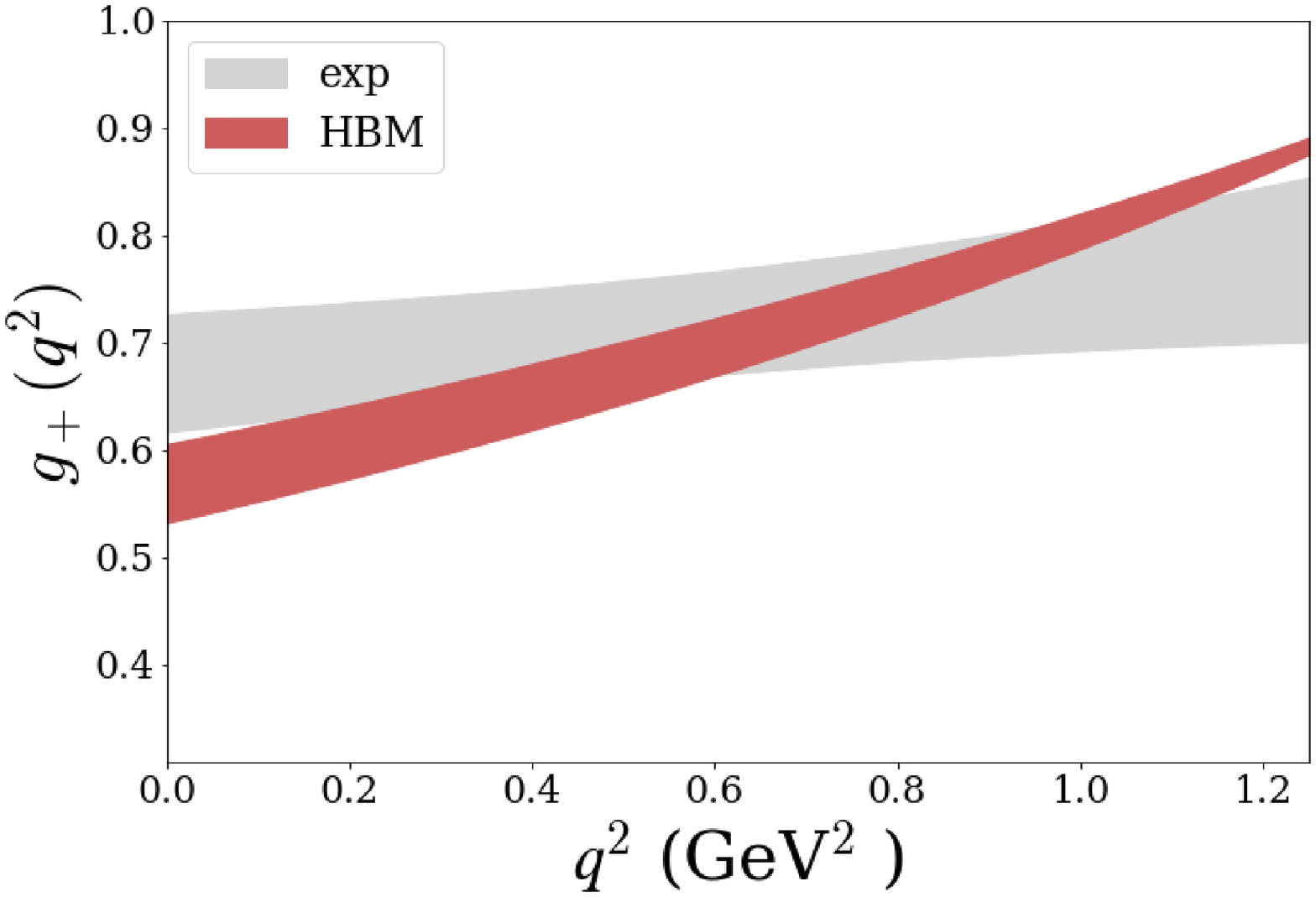} 
	\includegraphics[width=.4\textwidth]{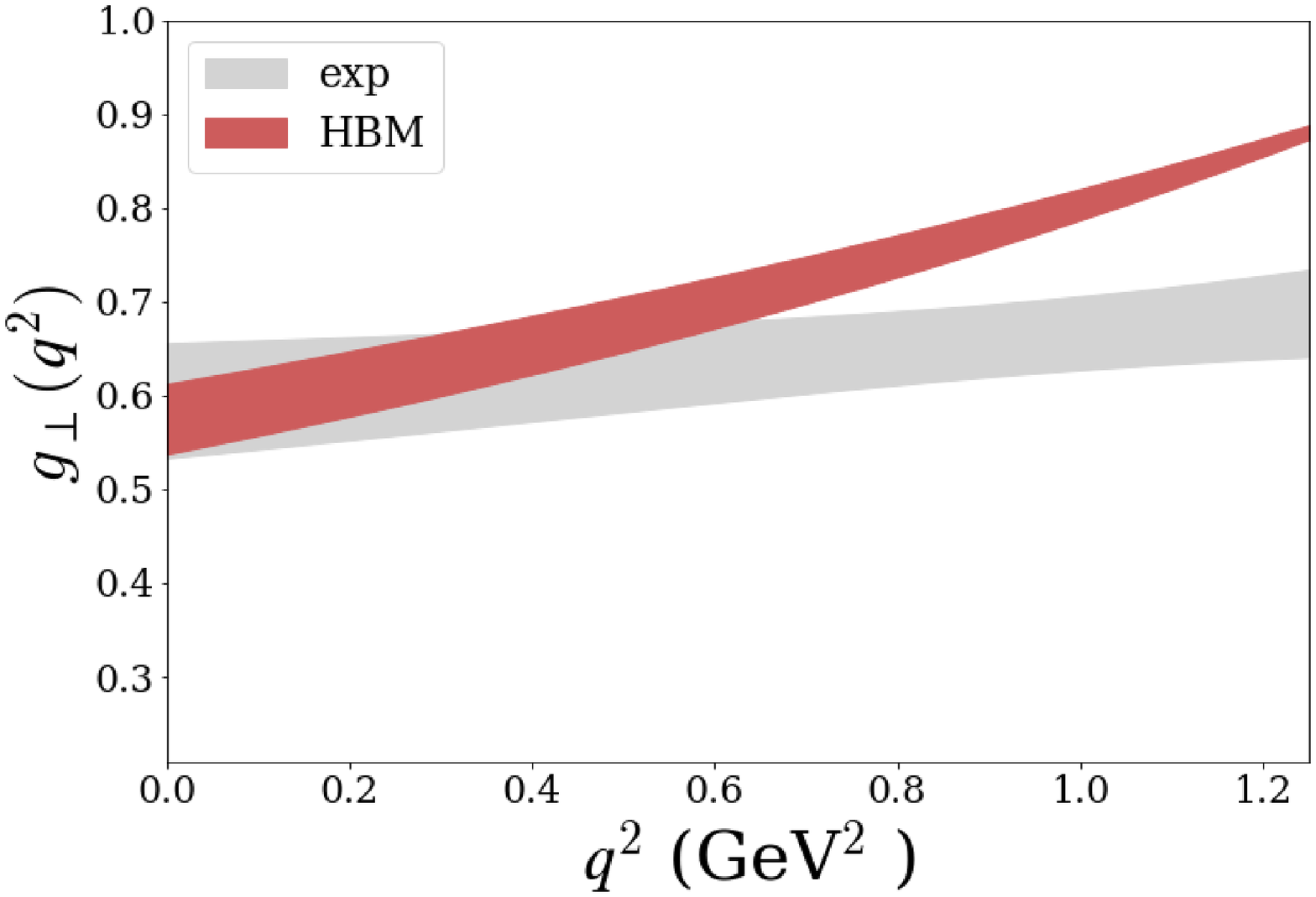}
	\caption{ The form factors obtained from HBM and the BESIII measurements. } 
	\label{cff}
\end{figure}

The results of HBM, LQCD and BESIII are given
in TABLE~\ref{form3} with $r_f= a^f_0 / a^{g_\perp}_0$, while the form factors from HBM and BESIII are drawn explicitly in FIG.~\ref{cff}. The HBM results of  $a_0^{g_\perp}, \alpha_1^{g_\perp}, \alpha_1^{g_+}$ lie between the ones of  LQCD and BESIII, whereas the signs of $\alpha_1^{f_\perp, f_+}$ are barely determined in the HBM.  Notice that $r_{g_+} =1 $ holds automatically from the definition of the form factors, which is also found explicitly in the experiments. FIG.~\ref{cff} shows that the main deviations between the results of HBM and BESIII  occur in $f_\perp$ at the high $q^2$ region.  

\begin{table}
\caption{Form factors compared with LQCD and experiments. }
\label{form3}
\begin{tabular}{lcccccccc}
\hline
\hline
&$a_0^{g\perp}$&$\alpha_1^{g_\perp}$&$\alpha_1^{g_+}$&$\alpha_1^{f_\perp}$&$\alpha_1^{f_+}$&$r_{f_+}$&$r_{f_\perp}$&$r_{g_+}$\\
\hline
LQCD &$0.68(2)$ &$-2.82(49)$&$-3.61(31)$&$-2.52(84)$&$-3.58(55)$&$1.19(3)$&$1.92(7)$&1\\
HBM  &$0.63(1)$& $-0.91(60)$  & $-1.05 (58)$&$0.16(66)$&$-0.29(72)$&$ 1.03(1)$&$1.51(5)$&1\\
Exp & $0.54(4)$&
\multicolumn{2}{c}{
$1.43(209)$}&\multicolumn{2}{c}{
$-8.15(158)$}  &$1.75(32)$&$3.62(65)$&$1.13(13)$
\\
\hline
\hline
\end{tabular}
\end{table}

	Some of the decay observables ${\cal X}_i$ have already been studied in the literature,  bearing different names. The polarization asymmetries  are
	defined as 
	\begin{eqnarray}\label{FB0}
		\alpha_{\Lambda} = \frac{\Gamma^\ell(\lambda_q = 1/2) -\Gamma^\ell(\lambda_q = - 1/2)  }{\Gamma^\ell(\lambda_q = 1/2) + \Gamma^\ell(\lambda_q = - 1/2)}  \,,
	\end{eqnarray}
which are extracted by the cascade decays of $\Lambda$ in the experiments, given as  
	\begin{equation}
		\alpha_\Lambda = \frac{2}{\alpha  } \left(  \int ^1_0 - \int^0 _{-1} \right) \Gamma^\ell_{\cos \theta_s }  d\cos \theta_s = \langle {\cal X}_9(\ell) \rangle \,. 
	\end{equation}
	Likewise, the forward-backward asymmetries are 
	\begin{eqnarray}\label{FB1}
		\alpha_{\ell} = \left(  \int ^1_0 - \int^0 _{-1} \right) \Gamma^\ell_{\cos \theta_\ell }  d\cos \theta_\ell  = \frac{1}{2}\langle {\cal X}_3(\ell) \rangle   \,,
	\end{eqnarray}
	describing the  distributions in  $W^{+*}\to \ell^+\nu$, while
	the up down asymmetries are 
	\begin{eqnarray}\label{FB}
		\alpha_{c}  = \frac{2}{P_b} \left(  \int ^1_0 - \int^0 _{-1} \right) \Gamma^\ell_{\cos \theta_c }  d\cos \theta_c  = \langle {\cal X}_4(\ell) \rangle  \,.
	\end{eqnarray}
	Note that we have adopted the shorthand  notations of
$
		\Gamma^\ell_{\Omega}  = \partial \Gamma ^\ell /( \Gamma^\ell \partial \vec{\Omega})   \,.
$

		\begin{table}
		\caption{  Results of $\alpha_\Lambda( \Lambda_c^+ \to \Lambda \ell^+\nu_{\ell})$ in units of $\%$.}
		\label{an}
		\begin{tabular}{l|cccc}
			\hline
			\hline
			&\multicolumn{1}{c}{$\alpha_\Lambda(\Lambda_c^+ \to \Lambda e^+\nu_e) $}
			&\multicolumn{1}{c}{$\alpha_\Lambda(\Lambda_c^+ \to \Lambda {\mu}^+\nu_{\mu}) $ }
			\\
			\hline
			HBM&$  -82.6(0)  $&$ -82.3(0) $\\
			LQCD~\cite{Meinel:2016dqj} &$ -87.4(10)$ & $-87.2(10)$\\
			Data~\cite{BESIII:2022ysa} &$ -86(4) $&$ - $\\
			RQM~\cite{Faustov:2019ddj} &$ -86$&$ - $\\
			LFQM ~\cite{Geng:2020gjh}&$ -97(3) $&$ -98(2) $\\
			LFQM ~\cite{Li:2021qod}&$ -87(9) $&$ -87(9) $\\
					SU(3)~\cite{Geng:2019bfz} &$ -86 (4) $&$ - $\\
			\hline
			\hline
			
		\end{tabular}
	\end{table}

	We compare $\alpha_\Lambda$ with those in the literature in TABLE~\ref{an}. In all the quark models, $\alpha_{\Lambda}$ depend little on the lepton flavors.  Except for those in Ref.~\cite{Geng:2020gjh}, $\alpha_{\Lambda}$ are consistent with the experimental values. 
 In contrast to other approaches, the predicted values of $\alpha_\Lambda$ in the HBM have little uncertainties  due to the correlations.
However, they are smaller than those in the LQCD. The future experiments on $\alpha_\Lambda$ shall be able to clarify the issue.

	\subsection{Observables in $\Lambda_c^+ \to n \ell^+ \nu_\ell $}
As the cascade decays of the neutron can not be observed in the experiments, 
there are only eight possible decay observables  in $\Lambda_c^+ \to n e^+ \nu$ , in which  five  of them require $P_b\neq 0$ for 
the measurements, given as 
	\begin{eqnarray}\label{exp}
	&& {\cal D}(q^2, \vec{\Omega})\propto 1 +  {\cal X}_2P_2+ {\cal X}_3 \cos \theta_\ell+ P_b \Big ( {\cal X}_4 \cos \theta_c + {\cal X}_5  \cos \theta_c P_2 + {\cal X}_6  \cos \theta_c \cos \theta_\ell \nonumber \\ 
	&& + \text{Re}( {\cal X}_7 e^{i \phi_{\ell}})\sin \theta_c \sin \theta_\ell + \text{Re} ({\cal X}_8e^{i \phi_{\ell}}) \sin \theta_c \sin \theta_\ell \cos \theta_\ell \Big )\,.
\end{eqnarray}
The discussions are parallel to $\Lambda_c^+\to \Lambda  \ell^+ \nu _\ell$. 
We list out the branching fractions and  decay observables in TABLE.~\ref{bn} and TABLE~\ref{n29}, respectively.
The branching fractions in the HBM  are  compatible with those in the LQCD~\cite{Meinel:2017ggx} and Ref.~\cite{Geng:2020gjh}, but twice larger than the results in Ref.~\cite{Zhao:2018zcb} . On the other hand,
the $SU(3)_F$ symmetry predicts  relatively large branching fractions comparing to the others. 

\begin{table}[t]
	\caption{  The branching fractions of $\Lambda_c^+ \to n \ell^+\nu_\ell$ in units of $\%$. }
	\label{bn}
	\begin{tabular}{l|cccc}
		\hline
		\hline
		&\multicolumn{1}{c}{${\cal B}(\Lambda_c^+ \to n e^+\nu_e )$}
		&\multicolumn{1}{c}{${\cal B}(\Lambda_c^+ \to n {\mu}^+\nu_{\mu}) $ }\\
		\hline
	HBM&$ 0.40(4)  $&$ 0.40 (4) $\\
				RQM~\cite{Faustov:2019ddj}&$ 0.268 $&$0.262 $\\
		LQCD~\cite{Meinel:2017ggx} &$ 0.410  (29) $&$ 0.400  ( 29)$\\
		LFQM~\cite{Zhao:2018zcb}&$0.201 $&$ - $\\
		LFQM~\cite{Geng:2020gjh} &$ 0.36 (15) $&$ 0.34 (15) $\\
	$	SU(3)_F$~\cite{He:2021qnc} &$0.520 (46) $&$0.506 (45) $\\
		\hline
		\hline
		
	\end{tabular}
\end{table}

\begin{table}[b]
	\caption{ The integrated observables in units of $\%$. }
	\label{n29}
	\begin{tabular}{ll|rrrrrr}
		\hline
		\hline
		\multirow{2}{*}{$i$}&\multirow{2}{*}{ {\centering ${\cal X}_i(\Lambda_c^+ \to n \ell^+ \nu) $  }}
		&\multicolumn{2}{c}{HBM}
		&\multicolumn{2}{c}{LQCD}\\
		&&$\langle {\cal X}_i(e)  \rangle$&$\langle {\cal X}_i(\mu )  \rangle$&$\langle {\cal X}_i(e)  \rangle$&$\langle {\cal X}_i(\mu )  \rangle$ \\
		\hline
		$2~$ & $(2\delta_\ell-1)(|a|^2 - \frac{1}{2} |b|^2)$  & $-42.3(13)$& $-36.5(10)$& $-43.4(16)$& $-37.2(15)$& \\
		$3$ & $ - 6\delta_\ell\left(\text{Re}\left( a_+t_+^*\right) +\text{Re} \left( a_-t_-^*\right) \right )+ \frac{3}{2}|b_\Delta|^2$  & $36.4(14)$& $30.0(17)$& $41.1(18)$& $34.4(18)$& \\
		$4$ & $(\delta_\ell + 1)(|a_\Delta|^2 + |b_\Delta|^2) + 3\delta_\ell |t_\Delta|^2$  & $-34.3(19)$& $-34.1(18)$& $-32.8(21)$& $-32.5(21)$& \\
		$5$ & $(2\delta_\ell-1)(|a_\Delta|^2 - \frac{1}{2}|b_\Delta|^2)$  & $70.7(4)$& $64.0(2)$& $73.9(13)$& $66.9(12)$& \\
		$6$ & $- 6\delta_\ell \left[\text{Re}(a_+t_+^*) - \text{Re}(a_-t_-^*)\right] - \frac{3}{2}|b_\Delta|^2$  & $-36.4(14)$& $-29.8(17)$& $-41.1(18)$& $-34.4(18)$& \\
		$7$ & $\frac{3}{\sqrt{2}}\left[ 2 \delta_\ell (t_-b_+^* - b_-t_+^*) + (a_-b_+^* + b_-a_+^*)\right]$  & $-91.7(9)$& $-91.9(9)$& $-92.1(7)$& $-92.3(7)$& \\
		$8$ & $ \frac{3}{\sqrt{2}} (2\delta_\ell- 1) (b_-a_+^* - a_-b_+^*) $  & $-78.7(7)$& $-74.7(8)$& $-83.2(15)$& $-79.0(14)$& \\
		\hline
		\hline
	\end{tabular}
\end{table}

\begin{figure}[b]
	\includegraphics[width=.3\textwidth]{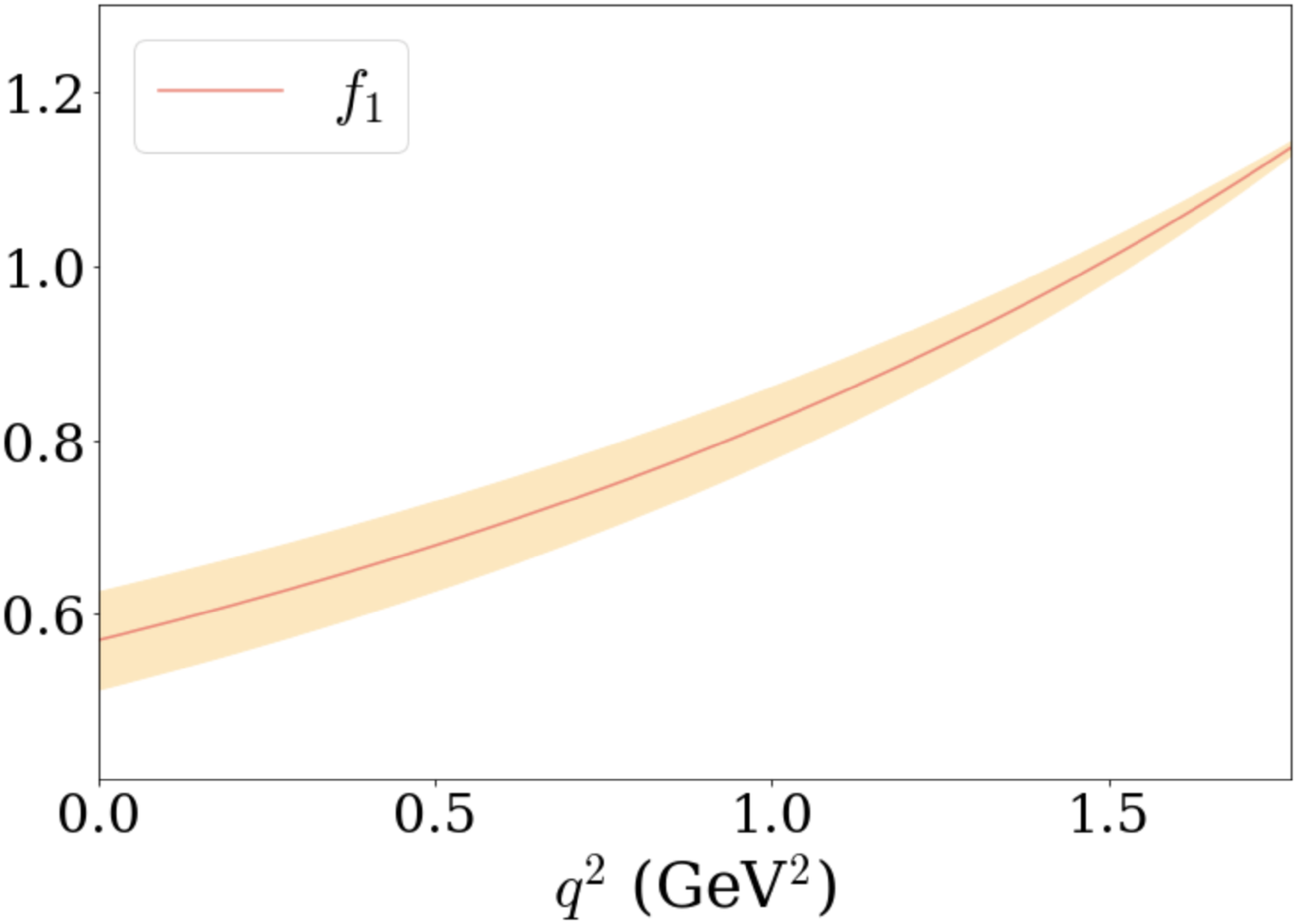}  
	\includegraphics[width=.3\textwidth]{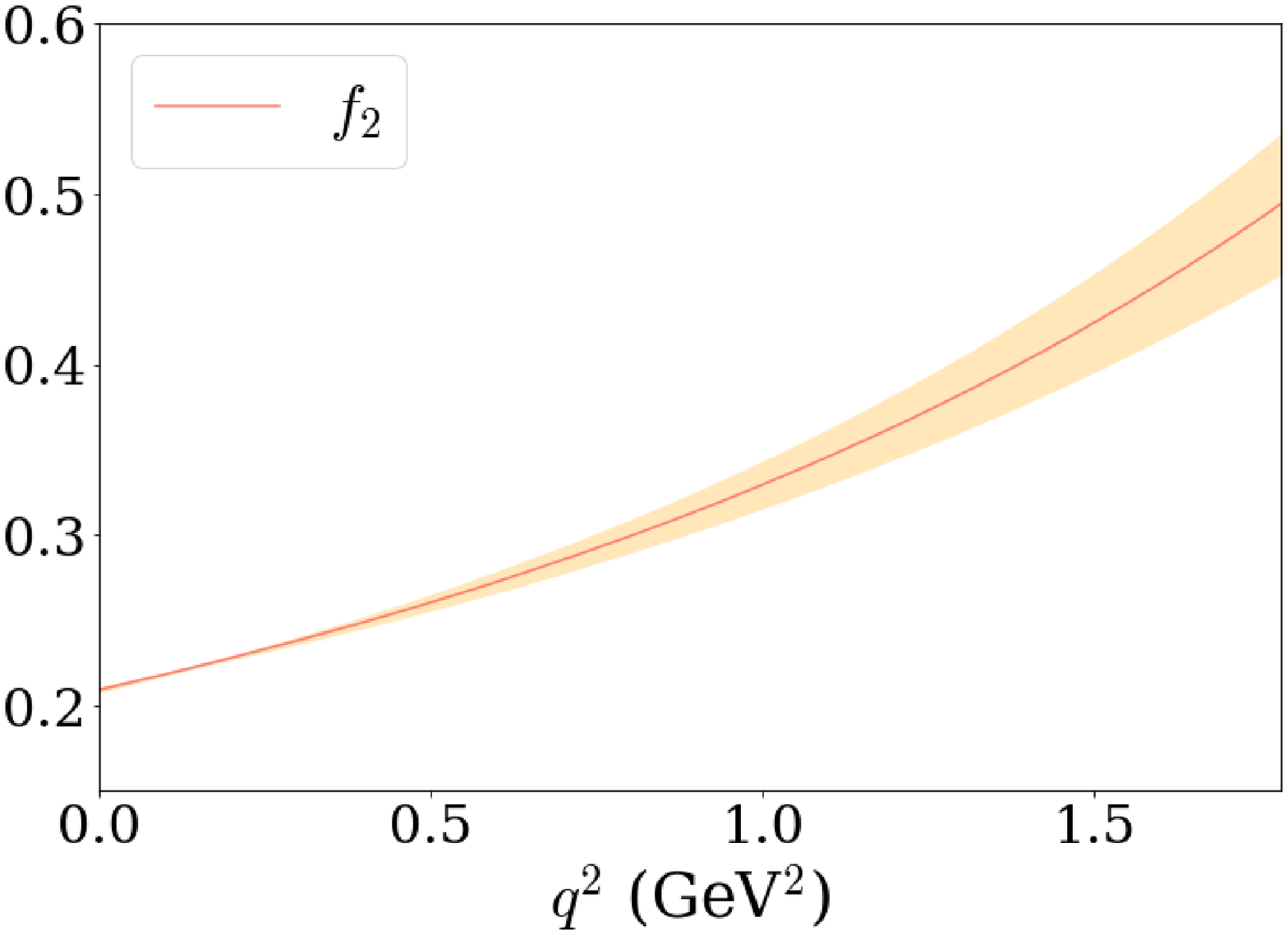} 
	\includegraphics[width=.3\textwidth]{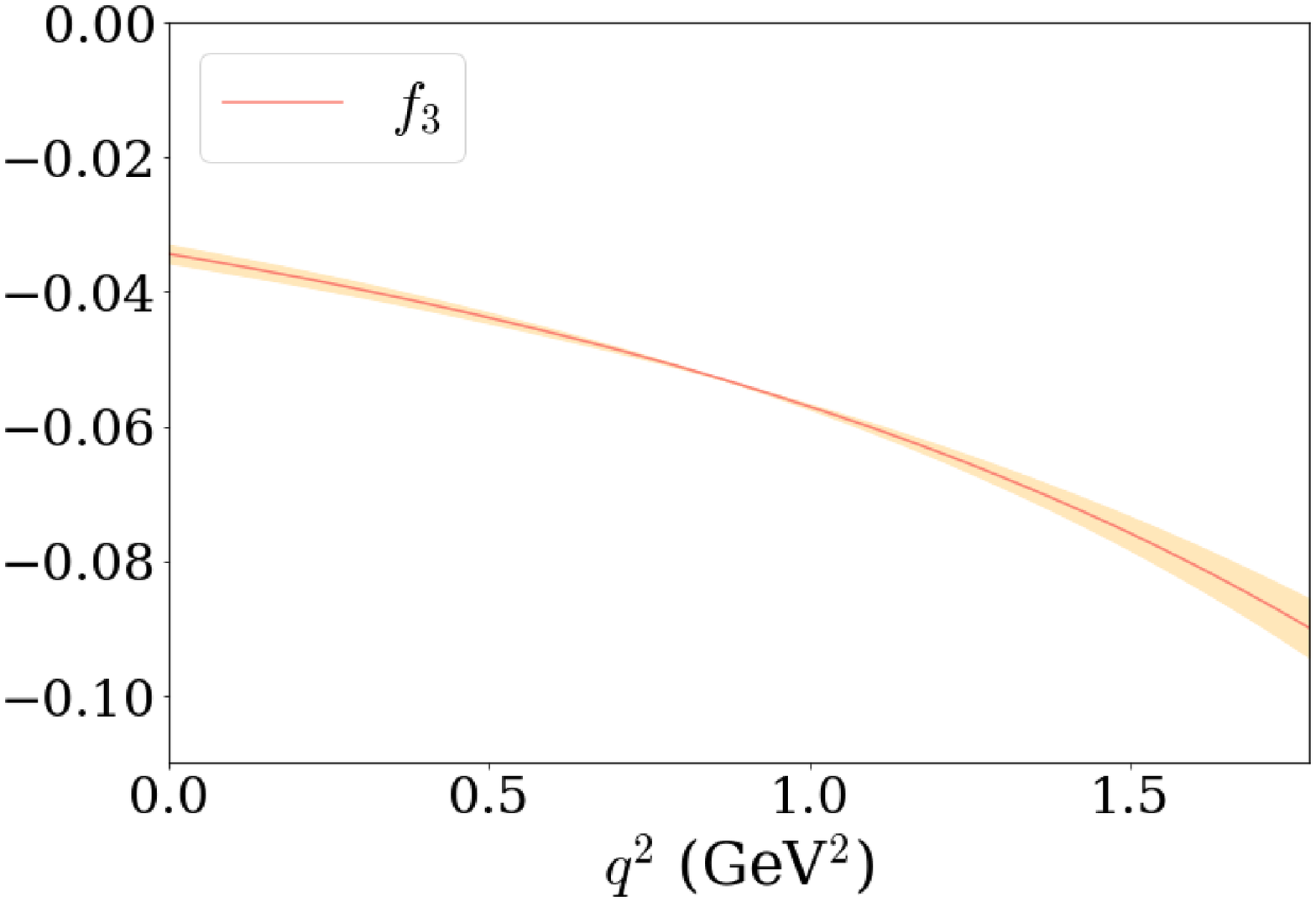} 
	\includegraphics[width=.3\textwidth]{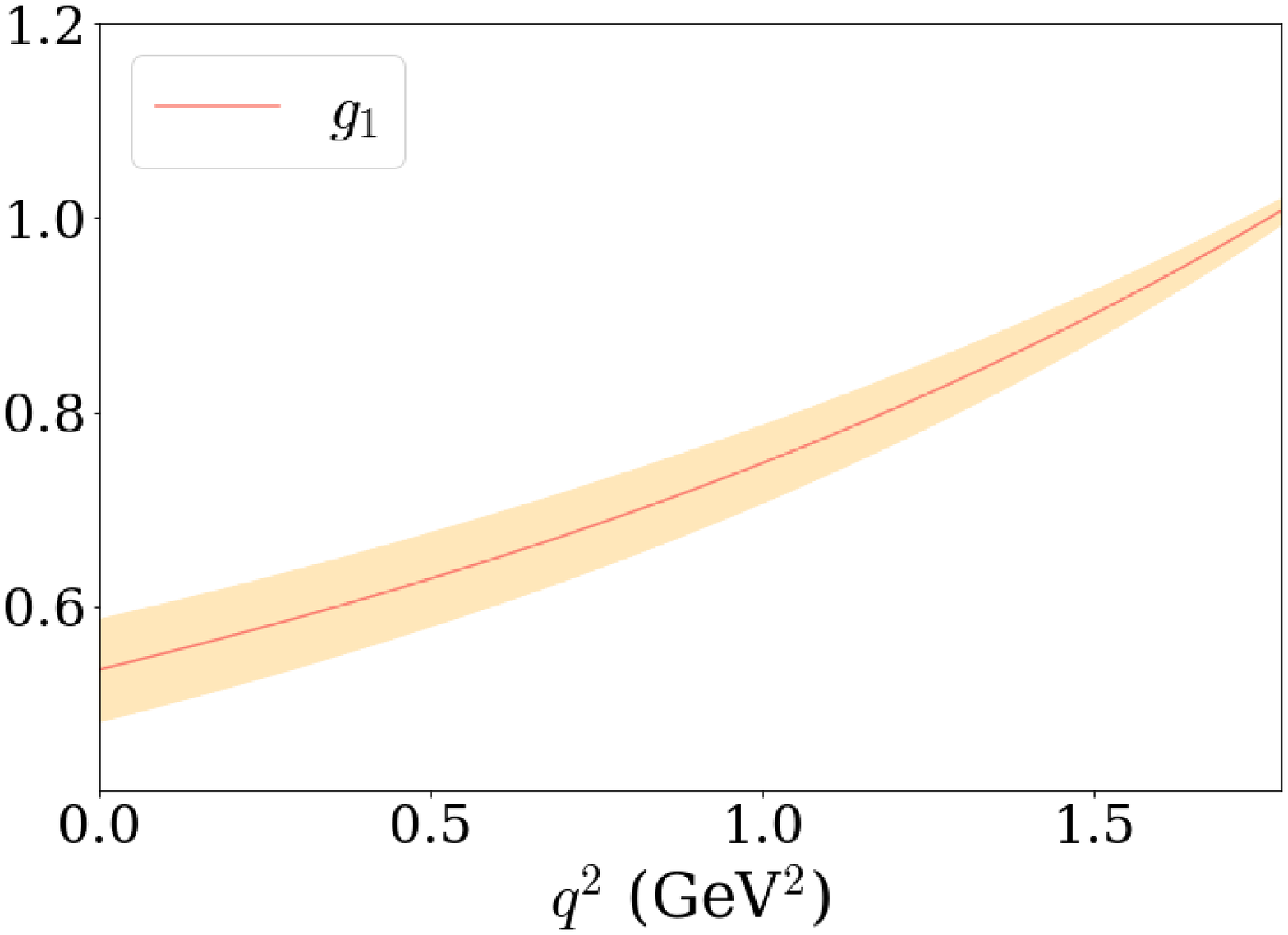}
	\includegraphics[width=.3\textwidth]{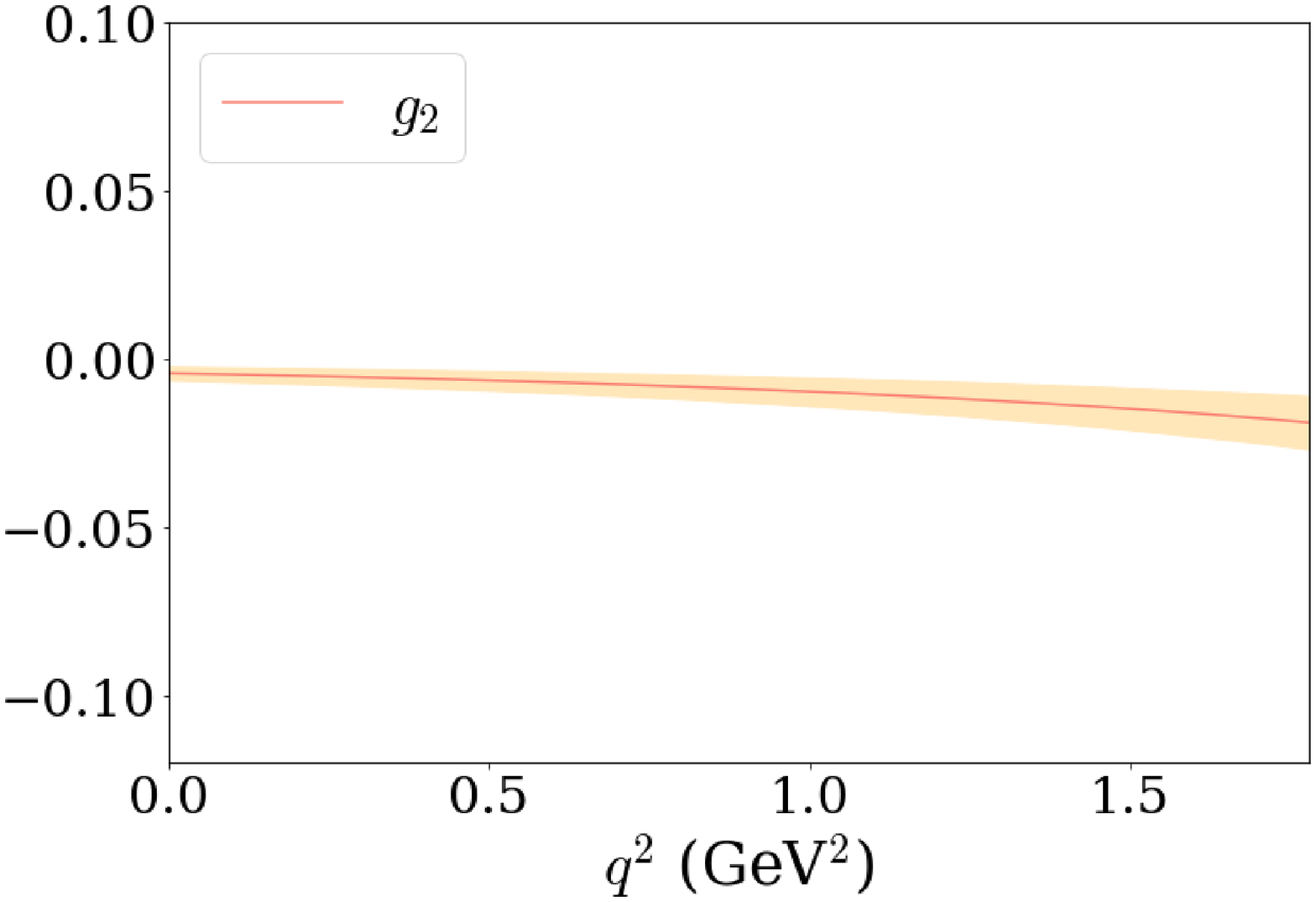} 
	\includegraphics[width=.3\textwidth]{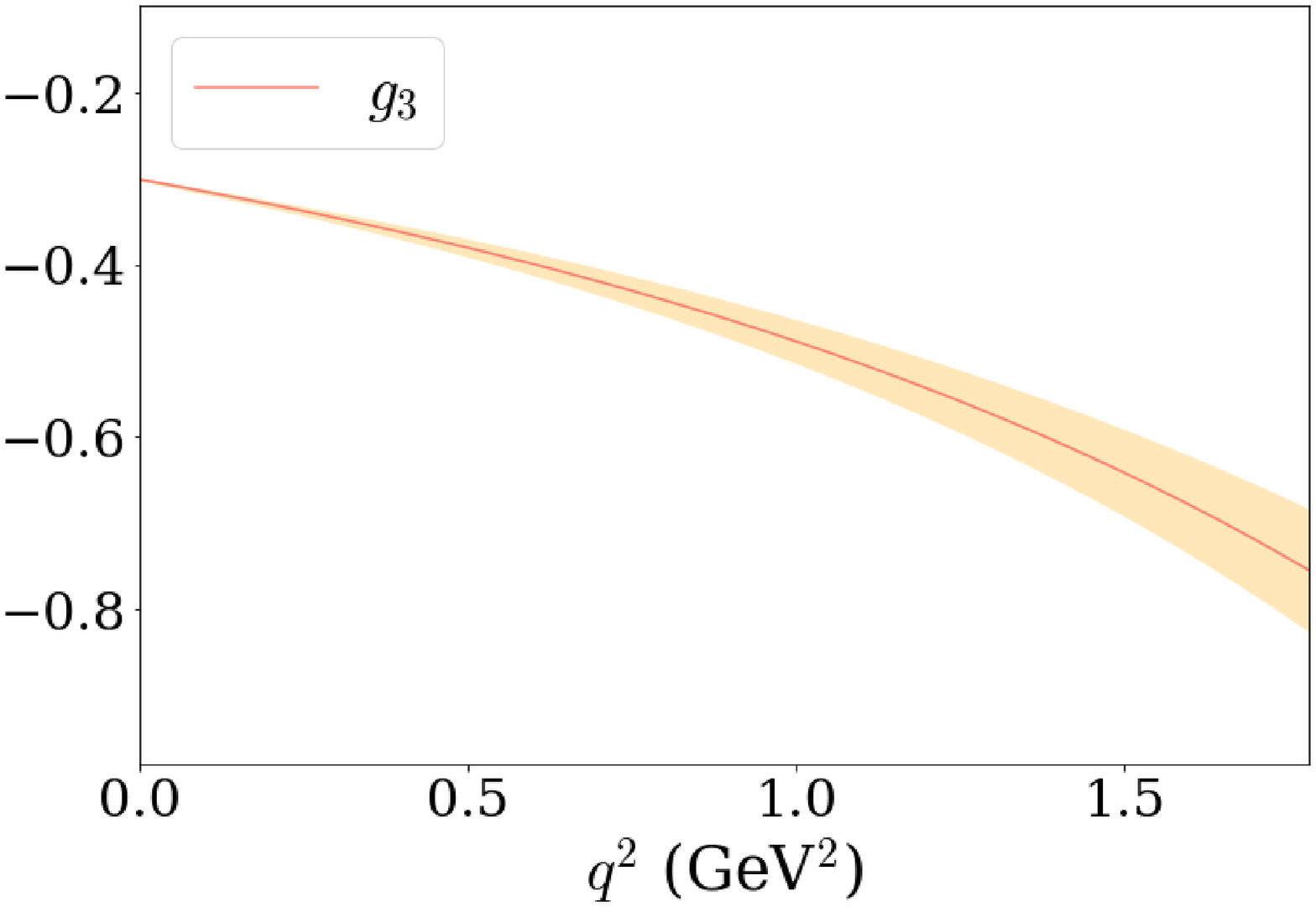}
	\caption{ The $q^2$ dependencies of the form factors in $\Lambda_c^+ \to n$ from HBM, where the bands represent the uncertainties from the model calculations. } 
	\label{formfactorsnfigure}
\end{figure}

The calculated values of $\langle {\cal X}_i(\ell)\rangle$ show consistencies between the HBM and LQCD. Nevertheless, ${\cal X}_{4\text{-}8}$ require $\Lambda_c^+$ to be polarized for  measurements, imposing  difficulties in the experiments.
The form factors of $\Lambda_c^+ \to n$ are given in TABLE.~\ref{form2} and FIG.~\ref{formfactorsnfigure}. We see that  once again the HBM underestimates $f_2$ by a factors of two third, whereas the others are compatible with the LQCD. 

\begin{table}
	\caption{ The form factors of $\Lambda_c^+ \to n$ at $q^2=0$.}
	\label{form2}
	\begin{tabular}{lcccccccc}
		\hline
		\hline
		&\multicolumn{1}{c}{$f_1(0) $ }
		&\multicolumn{1}{c}{$f_2(0)$}
		&\multicolumn{1}{c}{$g_1(0)$ }
		&\multicolumn{1}{c}{$g_2(0)$ }
		\\
		\hline 
	HBM &$ 0.570(56)  $&$ 0.210 (1) $&$ 0.526(50)  $&$ 0.015 (2) $\\
		LQCD~\cite{Meinel:2017ggx}  &$ 0.672 (39) $&$ 0.321 (38) $&$ 0.602 (31) $&$ -0.003 (52)$\\
		\hline
		\hline
	\end{tabular}
\end{table}

	\section{TIME REVERSAL ASYMMETRIES AND NEW PHYSICS}
	
In general, contaminated by the hadron uncertainties, it is often difficult to draw a sharp conclusion on whether NP is needed to explain the experimental data. 
Nevertheless,   NP can generate  clear signals of T asymmetries, which vanish in the SM and are equivalent to the CP asymmetries based on the CPT theorem. A great advantage of T violation is that  a strong phase is not necessary. The simplest T violating observables are studied in Ref.~\cite{Timereversal}, 
and we highlight some the results here.

Two of the simplest T asymmetries  in $\Lambda_c^+ \to \Lambda(\to p\pi^-) \ell^+ \nu_\ell$ are defined as 
\begin{eqnarray}\label{Time}
{\cal T}_\ell  	&=&\frac{1}{P_b }\left(  \int ^\pi _0 - \int_\pi ^ {2\pi } \right) \Gamma^\ell_{\phi_\ell}  d \phi_\ell   \quad = - \frac{\pi^2 }{8\Gamma^\ell } \int_{M_\ell^2}^{M_-^2}  \zeta {\bf \text{Im}}( {\cal X}_7 )  dq^2 \,,\nonumber\\
{\cal T}_s  &=& \frac{1}{\alpha }\left[ \left(  \int ^\pi _0 - \int_\pi  ^{2\pi } \right)d \Phi \right]  \left[ \left(  \int _0^{1}  - \int^0   _{-1 } \right)d \cos \theta_s   \right]   \Gamma^\ell_{\phi_\ell, \cos \theta_s }\nonumber\\
&=&	- \frac{2}{3 \pi \Gamma^\ell } \int_{M_\ell^2}^{M_-^2} \zeta {\bf \text{Im}}( {\cal X}_{12} )  dq^2 \,,
	\end{eqnarray}
which are closely related to the azimuthal angles. The reason is  that  they correspond to the triple product asymmetries of the three-momenta in the final states.  Keeping in mind  $\xi_\pm$ are real in the SM, it is easy to see that  nonvanishing ${\cal T}_{\ell,s}$  require NP beyond the SM. 

As an illustration, we explore NP with the effective Hamiltonian  as 
\begin{equation}
{\cal H}^{\text{NP}}_{eff} = \frac{G_F}{\sqrt{2}} V_{cs} \left[C_L \overline{s} \gamma_\mu (1-\gamma_5) c + C_R \overline{s} \gamma_\mu (1+\gamma_5) c \right]   \overline{v} \gamma_{\mu} (1-\gamma_5)u  \,.
\end{equation}
To the first order, we find that~\cite{Timereversal}
\begin{eqnarray}
&&\mathcal{T}_{\ell}=-\frac{3 \pi^2}{8 \sqrt{2}}\operatorname{Im}\left(C_R\right){\cal Y}_{\ell} \,,~~~
\mathcal{T}_{s}=-\frac{\sqrt{2}}{\pi }\operatorname{Im}\left(C_R\right){\cal Y}_{s}  \,,
	\end{eqnarray}
with
\begin{eqnarray}\label{Y}
		{\cal Y}_{\ell } &=& \frac{1 }{ \Gamma^\ell } \int_{M_\ell^2}^{M_-^2} 2 \zeta \left(  a_+ b_+ - a_- b_-\right)dq^2 \,,\nonumber\\
		{\cal Y}_{s} &=& \frac{1 }{ \Gamma^\ell } \int_{M_\ell^2}^{M_-^2} 2 \zeta \left(  a_+ b_- - a_- b_+\right)dq^2 \,,
	\end{eqnarray}
where 
$C_{L,R}$ are the Wilson coefficients from NP, and 
$\xi_ \pm$ in Eq.~\eqref{Y} are calculated within the SM. 
In practice, $C_L$ can be absorbed by redefining $V_{cs}$, so  their effects  vanish in the first order. 
We only considered NP with the left-handed neutrinos and leptons, as the right-handed ones are suppressed by the lepton masses~\cite{Timereversal}.

The values of ${\cal Y}_{\ell,s}$ can be viewed as the sensitivity coefficients of NP. 
We take Im$(C_R)= (0.1 , 0.2)$ as examples in TABLE~\ref{NewTable2}. We see  that   ${\cal T}_\ell$ can be as large as $10\%$, but they require $\Lambda_c^+$ to be polarized for an experimental measurement. On the other hand, though ${\cal T}_c$ can be probed with unpolarized $\Lambda_c^+$, their values are twice smaller than ${\cal T}_\ell$, making them hard to be observed. Finally, we emphasize that the T asymmetry  does {\it not} require a comparison with the charge conjugate as  the strong phase is irrelevant. 
	\begin{table}
		\caption{The time-reversal asymmetries versus Im$(C_R)$ from NP. }
		\label{NewTable2}
		\begin{tabular}{lcccccc}
			\hline
			\hline
&\multicolumn{2}{c}{Im$(C_R)=0.1$ } & 
\multicolumn{2}{c}{Im$(C_R)=0.2$ }
\\
 &$\ell^+ = e^+$ &$\ell^+ = \mu ^+$ &$\ell^+ = e^+$ &$\ell^+ = \mu ^+$\\
			\hline
	${\cal T}_\ell$~~&$0.062(2)$ &$ 0.060(2)$ & $0.124 (4)$ & $0.120(4) $ \\
	${\cal T}_s$ &$0.033(3)$ &$ 0.032(3)$& $0.066(1)$ &$ 0.064(1)$\\
			\hline
			\hline
			
		\end{tabular}
	\end{table}

\section{Summary}

We have given a systematical study on all the possible observables in $\Lambda_c^+$ semileptonic decays, including the effects of NP.
The model independent angular distributions with polarized $\Lambda_c^+$ have been presented. 
We have found that the BM underestimate $f_2$ by a factor of two third,  where the same underestimations were also found in the magnetic dipole moments of the octet baryons. 
  The branching fractions  and polarization asymmetries have been found to be 	
${\cal B}(\Lambda_c^+ \to \Lambda e^+ \nu _e, \Lambda \mu^+ \nu_\mu, n \ell ^+ \nu_\ell ) = (3.78 \pm 0.25, 3.67\pm 0.23, 0.40\pm 0.04 )\%$,and $\alpha_\Lambda (\Lambda_c^+ \to \Lambda e^+ \nu_e, \Lambda \mu^+ \nu _\mu ) =(-82.6, -82.3) \%$ from the HBM, 
and $\alpha_\Lambda (\Lambda_c^+ \to \Lambda e^+ \nu_e, \Lambda \mu^+ \nu _\mu ) =(-87.4\pm 1.0, -87.2\pm 1.0) \%$ from the LQCD, 
which can be tested in the near future at BESIII and Belle II.  
In particular,  ${\cal B}(\Lambda_c^+ \to \Lambda e^+ \nu _e, \Lambda \mu^+ \nu_\mu)$ and $\alpha_\Lambda (\Lambda_c^+ \to \Lambda e^+ \nu_e)$ are in good agreement with the current experimental data~\cite{BESIII:2016ffj,BESIII:2022ysa}.
All the angular observables in the SM have been computed by both the HBM and LQCD.
Most of the results from the two approaches  have been shown to be consistent.
The effects of NP on the T asymmetries have been explored, found to be  ${\cal O}(10\%)$ for Im$(C_R)=0.2$.  We stress that nonzero values of T asymmetries in the  experiments will be a  smoking gun of NP.

	\appendix 
	\section{Baryon wave functions}
For completeness, we display the wave functions of the low-lying baryons in the HBM. The wave functions of the octet baryons 
associated with the isospin of $I=1/2$  are  given as 
	\begin{eqnarray}
		&&	|n, \updownarrow\rangle = \int\frac{1}{2\sqrt{3} } \epsilon^{\alpha \beta \gamma} d _{a\alpha}^\dagger(\vec{x}_1) u_{b\beta}^\dagger(\vec{x}_2) d_{c\gamma}^\dagger (\vec{x}_3) \Psi_{A_\updownarrow(dus)}^{abc} (\vec{x}_1,\vec{x}_2,\vec{x}_3) [d^3  \vec{x}] | 0\rangle\,,\nonumber\\
		&&	|p, \updownarrow\rangle = \int\frac{1}{2\sqrt{3} } \epsilon^{\alpha \beta \gamma} d _{a\alpha}^\dagger(\vec{x}_1) u_{b\beta}^\dagger(\vec{x}_2) u_{c\gamma}^\dagger (\vec{x}_3) \Psi_{A_\updownarrow(dus)}^{abc} (\vec{x}_1,\vec{x}_2,\vec{x}_3) [d^3  \vec{x}] | 0\rangle\,,\nonumber\\
		&&	|\Lambda, \updownarrow\rangle = \int\frac{1}{\sqrt{6} } \epsilon^{\alpha \beta \gamma} d _{a\alpha}^\dagger(\vec{x}_1) u_{b\beta}^\dagger(\vec{x}_2) s_{c\gamma}^\dagger (\vec{x}_3) \Psi_{A_\updownarrow(dus)}^{abc} (\vec{x}_1,\vec{x}_2,\vec{x}_3) [d^3  \vec{x}] | 0\rangle\,,\nonumber\\
		&&	|\Xi^-, \updownarrow\rangle = \int\frac{1}{2\sqrt{3} } \epsilon^{\alpha \beta \gamma} d _{a\alpha}^\dagger(\vec{x}_1) s_{b\beta}^\dagger(\vec{x}_2) s_{c\gamma}^\dagger (\vec{x}_3) \Psi_{A_\updownarrow(dus)}^{abc} (\vec{x}_1,\vec{x}_2,\vec{x}_3) [d^3  \vec{x}] | 0\rangle\,,\nonumber\\
		&&	|\Xi^0, \updownarrow\rangle = \int\frac{1}{2\sqrt{3} } \epsilon^{\alpha \beta \gamma} u _{a\alpha}^\dagger(\vec{x}_1) s_{b\beta}^\dagger(\vec{x}_2) s_{c\gamma}^\dagger (\vec{x}_3) \Psi_{A_\updownarrow(dus)}^{abc} (\vec{x}_1,\vec{x}_2,\vec{x}_3) [d^3  \vec{x}] | 0\rangle\,,
	\end{eqnarray}
	and the ones with $I=1$ read as  
	\begin{eqnarray}
		&&	|\Sigma^{+}  , \updownarrow\rangle = \int\frac{1}{2\sqrt{3} } \epsilon^{\alpha \beta \gamma} u _{a\alpha}^\dagger(\vec{x}_1) u_{b\beta}^\dagger(\vec{x}_2) s_{c\gamma}^\dagger (\vec{x}_3) \Psi^{abc}_{S_\updownarrow (uuc)} (\vec{x}_1,\vec{x}_2,\vec{x}_3) [d^3  \vec{x}] | 0\rangle\,,\nonumber\\
		&&	|\Sigma^{0}  , \updownarrow\rangle = \int\frac{1}{\sqrt{6} } \epsilon^{\alpha \beta \gamma} d_{a\alpha}^\dagger(\vec{x}_1) u_{b\beta}^\dagger(\vec{x}_2) s_{c\gamma}^\dagger (\vec{x}_3) \Psi^{abc}_{S_\updownarrow (uuc)} (\vec{x}_1,\vec{x}_2,\vec{x}_3) [d^3  \vec{x}] | 0\rangle\,,\nonumber\\
		&&	|\Sigma^{-}  , \updownarrow\rangle = \int\frac{1}{2\sqrt{3} } \epsilon^{\alpha \beta \gamma} d_{a\alpha}^\dagger(\vec{x}_1) d_{b\beta}^\dagger(\vec{x}_2) s_{c\gamma}^\dagger (\vec{x}_3) \Psi^{abc}_{S_\updownarrow (uuc)} (\vec{x}_1,\vec{x}_2,\vec{x}_3) [d^3  \vec{x}] | 0\rangle\,.
	\end{eqnarray}
	The spin-flavor-antisymmetric spatial distribution $\Psi_A$ is defined in Eq.~\eqref{xdelta}, and the symmetric one is given as 
	\begin{equation}
		\begin{aligned}
			\Psi^{abc} _{S_\uparrow(q_1q_2q_3)} & (\vec{x}_1,\vec{x}_2,\vec{x}_3 ) =  \frac{{\cal N}_{{\cal B}}}{\sqrt{6}} \int \left(   2\phi^a_{q_1\uparrow}(\vec{x_1}') \phi^b_{q_2\uparrow}(\vec{x_2}') \phi^c_{q_3\downarrow}(\vec{x_3}') \right. \\
			&\left. -\phi^a_{q_1\uparrow}(\vec{x_1} ' ) \phi^b_{q_2\downarrow}(\vec{x_2}') \phi^c_{q_3\uparrow}(\vec{x_3}') 
			- \phi^a_{q_1\downarrow}(\vec{x_1}') \phi^b_{q_2\uparrow}(\vec{x_2}') \phi^c_{q_3\uparrow}(\vec{x_3}') \right)d^3 \vec{x}_\Delta \,,\\
			\Psi^{abc} _{S_\downarrow(q_1q_2q_3)} & (\vec{x}_1,\vec{x}_2,\vec{x}_3 ) =  \frac{{\cal N}_{{\cal B}}}{\sqrt{6}} \int \left(  -  2\phi^a_{q_1\downarrow}(\vec{x_1}') \phi^b_{q_2\downarrow}(\vec{x_2}') \phi^c_{q_3\uparrow}(\vec{x_3}') \right. \\
			&\left. + \phi^a_{q_1\downarrow}(\vec{x_1} ' ) \phi^b_{q_2\uparrow}(\vec{x_2}') \phi^c_{q_3\downarrow}(\vec{x_3}') 
			+  \phi^a_{q_1\uparrow}(\vec{x_1}') \phi^b_{q_2\downarrow}(\vec{x_2}') \phi^c_{q_3\downarrow}(\vec{x_3}') \right)d^3 \vec{x}_\Delta \,,
		\end{aligned} 
	\end{equation}
	where $\vec{x}'_i = \vec{x}_i  - \vec{x}_\Delta$. 
	
	Similarly, the  heavy baryons with a single heavy quark are given  as 
	\begin{eqnarray}
		&&	|\Lambda_c^+, \updownarrow\rangle = \int\frac{1}{\sqrt{6} } \epsilon^{\alpha \beta \gamma} d _{a\alpha}^\dagger(\vec{x}_1) u_{b\beta}^\dagger(\vec{x}_2) c_{c\gamma}^\dagger (\vec{x}_3) \Psi_{A_\updownarrow(duc)}^{abc} (\vec{x}_1,\vec{x}_2,\vec{x}_3) [d^3  \vec{x}] | 0\rangle\,,\nonumber\\
		&&	|\Xi_c^+, \updownarrow\rangle = \int\frac{1}{\sqrt{6} } \epsilon^{\alpha \beta \gamma} u _{a\alpha}^\dagger(\vec{x}_1) s_{b\beta}^\dagger(\vec{x}_2) c_{c\gamma}^\dagger (\vec{x}_3) \Psi_{A_\updownarrow(usc)}^{abc} (\vec{x}_1,\vec{x}_2,\vec{x}_3) [d^3  \vec{x}] | 0\rangle\,,\nonumber\\
		&&	|\Xi_c^0, \updownarrow\rangle = \int\frac{1}{\sqrt{6} } \epsilon^{\alpha \beta \gamma} d _{a\alpha}^\dagger(\vec{x}_1) s_{b\beta}^\dagger(\vec{x}_2) c_{c\gamma}^\dagger (\vec{x}_3) \Psi_{A_\updownarrow(dsc)}^{abc} (\vec{x}_1,\vec{x}_2,\vec{x}_3) [d^3  \vec{x}] | 0\rangle\,,
	\end{eqnarray}
	for the antitriplet baryons, and 
	\begin{eqnarray}
		&&	|\Sigma_c^{  ++}  , \updownarrow\rangle = \int\frac{1}{2\sqrt{3} } \epsilon^{\alpha \beta \gamma} u _{a\alpha}^\dagger(\vec{x}_1) u_{b\beta}^\dagger(\vec{x}_2) c_{c\gamma}^\dagger (\vec{x}_3) \Psi^{abc}_{S_\updownarrow (uuc)} (\vec{x}_1,\vec{x}_2,\vec{x}_3) [d^3  \vec{x}] | 0\rangle\,,\nonumber\\
		&&	|\Sigma_c^{  +}  , \updownarrow \rangle = \int\frac{1}{\sqrt{6} } \epsilon^{\alpha \beta \gamma} d _{a\alpha}^\dagger(\vec{x}_1) u_{b\beta}^\dagger(\vec{x}_2) c_{c\gamma}^\dagger (\vec{x}_3) \Psi^{abc}_{S_\updownarrow (duc)} (\vec{x}_1,\vec{x}_2,\vec{x}_3) [d^3  \vec{x}] | 0\rangle\,,\nonumber\\
		&&	|\Sigma_c^{  0 }  , \updownarrow \rangle = \int\frac{1}{2 \sqrt{3} } \epsilon^{\alpha \beta \gamma} d _{a\alpha}^\dagger(\vec{x}_1) d_{b\beta}^\dagger(\vec{x}_2) c_{c\gamma}^\dagger (\vec{x}_3) \Psi^{abc}_{S_\updownarrow (ddc)} (\vec{x}_1,\vec{x}_2,\vec{x}_3) [d^3  \vec{x}] | 0\rangle\,,\nonumber\\
		&&	|\Xi_c^{ \prime +}  , \updownarrow \rangle = \int\frac{1}{\sqrt{6} } \epsilon^{\alpha \beta \gamma} u _{a\alpha}^\dagger(\vec{x}_1) s_{b\beta}^\dagger(\vec{x}_2) c_{c\gamma}^\dagger (\vec{x}_3) \Psi^{abc}_{S_\updownarrow (usc)} (\vec{x}_1,\vec{x}_2,\vec{x}_3) [d^3  \vec{x}] | 0\rangle\,,\nonumber\\
		&&	|\Xi_c^{\prime 0} , \updownarrow \rangle = \int\frac{1}{\sqrt{6} } \epsilon^{\alpha \beta \gamma} d _{a\alpha}^\dagger(\vec{x}_1) s_{b\beta}^\dagger(\vec{x}_2) c_{c\gamma}^\dagger (\vec{x}_3) \Psi^{abc}_{S_\updownarrow (dsc)} (\vec{x}_1,\vec{x}_2,\vec{x}_3) [d^3  \vec{x}] | 0\rangle\,,\nonumber\\
		&&	|\Omega_c^{ 0}, \updownarrow \rangle = \int\frac{1}{2 \sqrt{3} } \epsilon^{\alpha \beta \gamma} s _{a\alpha}^\dagger(\vec{x}_1) s_{b\beta}^\dagger(\vec{x}_2) c_{c\gamma}^\dagger (\vec{x}_3) \Psi^{abc}_{S_\updownarrow (ssc)} (\vec{x}_1,\vec{x}_2,\vec{x}_3) [d^3  \vec{x}] | 0\rangle\,,
	\end{eqnarray}
	for the sextet baryons. The bottom baryons are obtained directly by substituting  bottom quarks for the charmed quarks.

	On the other hand, the low-lying spin $3/2$ baryons are constructed by 
	\begin{equation}
		|{\cal B},J=\frac{3}{2}, J_z \rangle = \int\frac{ 1}{\sqrt{6S_{\cal B}!} } \epsilon^{\alpha \beta \gamma} q_{1 a\alpha}^\dagger(\vec{x}_1) q_{2b\beta}^\dagger(\vec{x}_2) q_{3c\gamma}^\dagger (\vec{x}_3) \Psi^{abc}_{T J_z (q_1q_2q_3)} (\vec{x}_1,\vec{x}_2,\vec{x}_3) [d^3  \vec{x}] | 0\rangle\,,
	\end{equation}
	where $S_{\cal B}$ is the number of the identical quark in the baryon, 
	$J_z$ the angular momentum in $\hat{z}$ direction, and 
	\begin{equation}\label{both}
		\begin{aligned}
			\Psi^{abc}_{T\frac{3}{2} (q_1q_2q_3)}&(\vec{x}_1,\vec{x}_2,\vec{x}_3) =  {\cal N}_{{\cal B}} \int   \phi^a_{q_1\uparrow}(\vec{x_1}') \phi^b_{q_2\uparrow}(\vec{x_2}') \phi^c_{q_3\uparrow}(\vec{x_3}')  d^3 \vec{x}_\Delta \,,\\
			\Psi^{abc}_{T\frac{1}{2} (q_1q_2q_3)}& (\vec{x}_1,\vec{x}_2,\vec{x}_3) =  \frac{ {\cal N}_{{\cal B}}}{\sqrt{3}} \int   \big( \phi^a_{q_1\downarrow}(\vec{x_1}') \phi^b_{q_2\uparrow}(\vec{x_2}') \phi^c_{q_3\uparrow}(\vec{x_3}')  d^3 \vec{x}_\Delta  \\
			&
			+ \phi^a_{q_1\uparrow}(\vec{x_1}') \phi^b_{q_2\downarrow}(\vec{x_2}') \phi^c_{q_3\uparrow}(\vec{x_3}') + \phi^a_{q_1\uparrow}(\vec{x_1}') \phi^b_{q_2\uparrow}(\vec{x_2}') \phi^c_{q_3\downarrow}(\vec{x_3}') \big)   d^3 \vec{x}_\Delta 
			\,.
		\end{aligned}
	\end{equation}
	The baryon wave functions with negative angular momenta can be obtained by flipping the spin directions in both sides of Eq.~\eqref{both}.

	\begin{acknowledgments}
The authors are indebted  to Stefan Meinel for the help  on the LQCD form factors.
		This work is supported in part by the National Key Research and Development Program of China under Grant No. 2020YFC2201501 and  the National Natural Science Foundation of China (NSFC) under Grant No. 12147103.
	\end{acknowledgments}

\end{document}